\documentclass[useAMS,usenatbib]{mn2e}

\title[HDF--N \& HDF--S Quantitative Morphological Analysis]
{Quantitative Morphological Analysis of the HDF--North and HDF-South. I. Early--
and Late--type Luminosity--Size Relations of galaxies out to
$z$$\sim$1}

\author[I. Trujillo and J.A.L. Aguerri]{I. Trujillo $^{1}$\thanks{E-mail:
trujillo@mpia-hd.mpg.de (IT); jalfonso@iac.es (JALA)} and J. A. L. 
Aguerri$^{2}$\footnotemark[1]\\
$^{1}$Max--Planck--Institut f\"ur Astronomie, K\"onigstuhl 17, D--69117
Heidelberg, Germany\\
$^{2}$Instituto de Astrof\'{\i}sica de Canarias, Calle V\'{\i}a L\'actea,
E-38200 La Laguna, Tenerife, Spain}

\begin{document}

\date{Accepted 0000 December 00. Received 0000 December 00; in original form
0000 October 00}

\pagerange{\pageref{firstpage}--\pageref{lastpage}} \pubyear{2002}

\maketitle

\label{firstpage}

\begin{abstract}

Based on  drizzled F606W and F814W images, we present  quantitative structural
parameters in the V--band rest-frame  for all  galaxies with $z$$<$1 and 
I$_{814}$(AB)$<$24.5 mag in the Hubble Deep Fields North and South. Our
structural parameters are based on a two--component surface brightness
distribution using a S\'ersic bulge  and an exponential disc. Detailed
simulations and comparisons with previous work are presented. The   
luminosity--size distribution of early--type galaxies is consistent with the
hypothesis that their structural properties  were already in place by $z\sim$1
and have evolved passively since then; early--type galaxies were $\sim$
1.35($\pm$0.1) mag brighter in rest--frame V--band luminosity at $z$$\sim$0.7
than now. Compared to present day late--type galaxies, those at $z$$\sim$0.7
with  L$_V$$>$0.2$\times$10$^{10}$ h$^{-2}$ L$_\odot$ show a moderate decrease
($\sim$30($\pm$10)\%) in  size (or interpreted differently, a decrease of
$\sim$0.77($\pm$0.30) mag in the central surface brightness) at a given
luminosity. Finally, we make a comparison of our results with the infall and
hierarchical models.

\end{abstract}


\section{Introduction}

The Hubble Deep Field (HDF)  North (Williams et al. 1996) and  South (Casertano
et al. 2000) are among the deepest views of the universe ever obtained.  These
observations were specially made to facilitate the study of galaxy evolution.
The  high resolution of the publically  available drizzled images
(0.04$''$/pix, i.e. $\sim$0.22 h$^{-1}$ kpc at $z$$\sim$1\footnote{We adopt a
$\Lambda$--cosmology throughout the paper with $\Omega_M$=0.3,
$\Omega_\Lambda$=0.7 and H$_0$=100h km s$^{-1}$ Mpc$^{-1}$.}) convert them into
an ideal laboratory for morphological evolutionary studies. 

Morphology plays a fundamental role in understanding  galaxy evolution  as
each galaxy type could evolve differently with $z$. To  facilitate the
comparison with the theoretical predictions, galaxy evolution studies must be
rooted in a quantitative basis. In addition,   quantitative morphology
enables one to study any possible evolution of the galaxy structural parameters
such as luminosity, size, etc., and their associated scaling relations with
regards to the values measured in nearby samples. 

In the  hierarchical galaxy formation framework, both disc and elliptical
galaxy properties are expected to  evolve at high--z. In this paradigm, it is
generally assumed that much of a galaxy's star formation takes place in a
galactic disc which formed by gas infall into dark matter halos (White \& Frenk
1991; Kauffmann et al. 1999; Somerville \& Primack 1999). According to Mo, Mau
\& White (1998) the size of the disc galaxies, at a fixed halo circular
velocity, must decrease with redshift as $R\propto$H$^{-1}$(z) where H(z) is
the Hubble parameter at redshift z. On the other hand, E/S0 galaxies are
expected to form via repeated merging (Toomre \& Toomre 1972). The formation
epoch of the E/S0 galaxies is sensitive to the assumed values of cosmological
parameters, occurring at higher redshift at lower  $\Omega_M$ (Kauffmann \&
Charlot 1998).

There are alternative theories of galaxy formation. One is that all galaxies,
including E/S0, formed very early via monolithic collapse of the gas at high
redshift (Eggen, Lynden--Bell \& Sandage 1962; Larson 1975; Chiosi \& Carraro
2002). According to this theory  E/S0 properties will be already in place at
high--z. For  disc galaxies, some ``backward'' approaches have been tried.
These approaches use the detailed studies of the profiles of the Milky way and
other nearby galaxies to propose radially dependent models of star formation
that could be used to make predictions about high--z disc evolution (Cay\'on,
Silk \& Charlot 1996; Bouwens, Cay\'on \& Silk 1997; Roche et al. 1998).
However, both the hierarchical and ``backward'' models predict relatively
similar amounts of evolution in disc global properties (size, luminosity, etc.)
to $z$$\sim$1 (Bouwens \& Silk 2002).

The field covered by both HDFs is not large compared to others works but they
are so deep that they are significantly less affected by the incompleteness
problem affecting other less deep samples. Discrepancies associated to
incompleteness can be found in the literature, for example, studying discs
galaxies, Schade et al. (1995, 1996) who used early ground-- and space--based
images of galaxies from the Canada--France Redshift Survey (CFRS) and Roche et
al. (1998) in the Medium Deep Survey found a net increase in the surface
brightness and a net decrease in the size. However, Simard et al. (1999), who
conducted a detailed analysis of the selection effects in the Deep
Extragalactic Evolutionary Probe (DEEP) survey found no evolution in  disc
properties up to $z$$\sim$1.

In the following work we conduct a detailed morphological analysis of all the
galaxies with z$<$1 in the HDF--North and South  on the drizzled F606W and
F814W images down to I$_{814}$(AB)$=$24.5 mag. Quantitative morphological
analysis can be conducted using non--model based techniques like the
concentration and asymmetry of the galactic light (Abraham et al. 1996).
However, these techniques have the disadvantage that they do not allow the
study of the different structural components of the galaxy separately, and
particularly,  the evolutionary analysis of the well known local scale relation
can not be addressed.

 Improving upon previous analyses, our structural parameters are based on a
two--component surface brightness distribution using a S\'ersic $r^{1/n}$ bulge
 (S\'ersic 1968) and an exponential disc. A previous detailed
morphological analysis of the HDF--N was carried by Marleau \& Simard
(1998, hereafter MS98). However, that work only considered a $r^{1/4}$ bulge,
instead of a free $r^{1/n}$. Our assumption of a $n$ free bulge is more in
agreement to what it is observed in the nearby universe (see, e.g. Graham 2001
and references therein). On the other hand, even though the nearly identical
quality of the HDF--S, a systematic quantitative morphological analysis of this
field have been not yet conducted and it is of great interest to compare
 the galaxy properties between these two twin fields.

In this paper we present the galaxy structural parameters of both HDF  fields
and compare our results to previous classification schemes. To assess the
degree of evolution, if any, the early-- and late--type luminosity--size
relations are compared with those observed in the nearby universe provided by
large local surveys such as the Sloan Digital Sky Survey (York et al. 2000). In
a second paper (Aguerri \& Trujillo 2004), we analyse the redshift evolution of
the structural components of the galaxies (bulge and disc) separately. The
outline of this paper is as follows: section 2 describes the characteristics of
the observed fields. Our quantitative morphological analysis algorithm is
explained in section 3. In section 4 we detail our object morphological
classification and in section 5 we analyse the accuracy of our analysis by
comparing our results with simulations and previous work. Section 6  deals with
the issue of selection effects. The luminosity--size relations of early-- and
late--type galaxies  are presented in section 7. We discuss our results in
section 8.

\section[]{Data}

 Each HDF image consists on a $\approx$ 2.5 $\times$ 2.5 arcmin$^2$  Wide Field
Planetary Camera 2 (WFPC2) pointing (Williams et al. 1996, HDF--N; Casertano et
al. 2000, HDF--S). The images were acquired through the broad band F300W,
F450W, F606W, and F814W filters. For the present work we have used the
publically available reduced F606W and F814W Version 2 Drizzled images  with  a
sampling of 0.04 arcsec/pixel and a total exposure time of  109050 s
(F606W;HDF--N), 123600 s (F814W;HDF--N), 97200 s (F606W;HDF--S)  and 112200 s
(F814W;HDF--S). The high sampling of these images avoids the undersampling
problems affecting other Hubble Space Telescope (HST) images. 

In order to assure a reliable estimation of the structural parameters (see the
simulations section) we have selected all the  I$_{814}$(AB)$<$24.5  galaxies
with z$<$1. In addition,  to maintain the analysis in the V--band rest--frame,
we have also studied the  I$_{814}$(AB)$<$24.5  galaxies with z$<$0.4 in the
F606W filter. The above restrictions  leave us with 123 objects in the HDF--N
and 95 in the HDF--S. The redshift and total magnitude information was obtained
from two photometric redshift catalogs: Fern\'andez--Soto, Lanzetta \& Yahil,
(1999; HDF--N) and Labb\'e et al. (2003; HDF--S). The HDF--N catalog  provides
aperture magnitudes which are then corrected for the effect of neighboring.
These magnitudes correspond to the SExtractor magnitudes named as ``best''. For
the HDF--S catalog, Labb\'e et al. provide with a ``color'' aperture (see their
sec. 5.2). This ``color'' aperture is  an isophotal aperture  determined by the
K$_s$--band detection isophote at the 5 $\sigma$ detection threshold when the
galaxies satisfy a  ``size'' criteria, and is a fixed aperture when the
galaxies are outside this ``size'' range. This ``color'' magnitude is finally
transformed to a total magnitude using the total magnitude in the K$_s$--band.
The total K$_s$--band magnitude is based on a Kron--like aperture (SEx AUTO)
derived from the K$_s$--detection image. When available, spectroscopic
redshifts  are used. This corresponds to 70\% of the galaxies in the HDF--N and
60\% in the HDF--S in the present work.

The number identifications of the objects correspond to the same numbers used
in the photometric redshift catalogs. 

\section{The Fitting Algorithm}

Our morphological analysis is based on the decomposition of the surface
brightness (SB) profile into a bulge and a disc component. The bulge component
is modeled by a S\'ersic (1968) profile of the form:
\begin{equation}
I(r)=I_{b}(0)e^{-b_n(r/r_{e,b})^{1/n}}
\end{equation}
 where $I_{b}(0)$ is the bulge central intensity, $r_{e,b}$ is the bulge
  semi--major
 effective radius and $n$ the S\'ersic shape parameter. The intensity at the
 bulge effective radius $r_{e,b}$ is given by $I_{e,b}$=$I_{b}(0)$$e^{-b_n}$. 
  The quantity b$_n$ is a
 function of the shape paramenter $n$, and is chosen so that $r_{e,b}$ encloses
 half of the total luminosity. The exact value is derived from
 $\Gamma(2n)=2\gamma(2n,b_n)$, where $\Gamma(a)$ and $\gamma(a,x)$ are the gamma
 and the incomplete gamma function (Abramowitz \& Stegun 1964, p. 260).
   The S\'ersic model contains the
 classical de Vaucouleurs profile when $n$ is equal to 4. The disc component is an exponential
 profile (Freeman 1970) given by:
\begin{equation}
I(r)=I_d(0)e^{-r/h}
\end{equation}
where $I_d(0)$ is the central intensity and $h$ is the  disc
semi--major scale--length. The set of free parameters is completed with the
ellipticities of the bulge $\epsilon_{b}$ and the disc $\epsilon_{d}$. In
addition, it is possible to determine from the above parameters
 the total magnitude:
\begin{equation}
I_{814}(model)=-2.5\log((F_b+F_d)/t)+C
\end{equation}
where $t$ is the exposure time, C is the zeropoint, $F_b$ is the bulge luminosity 
$F_b=I_b(0)r^2_{e,b}(1-\epsilon_{b})2\pi n\Gamma(2n)/b^{2n}_n$ and $F_d$ 
is the disc
luminosity $F_d=I_d(0)h^2(1-\epsilon_{d})2\pi$. We have used C=23.02 (F606W) and
C=22.09 (F814W) in the AB system.

We can also evaluate
the bulge--to--total luminosity ratio B/T:
\begin{equation}
\frac{B}{T}=\frac{F_b}
{F_b+F_d}
\end{equation}
and the global semi--major effective radius r$_e$ by solving the equation:
\begin{equation}
\frac{B/T}{1-B/T}
\biggr[1-\frac{2\gamma(2n,b_n(r_e/r_{e,b})^{1/n})}{\Gamma(2n)}\biggr]=
2\gamma(2,r_e/h)-1
\end{equation}

\begin{figure*}

\vskip 16.0cm
{\includegraphics{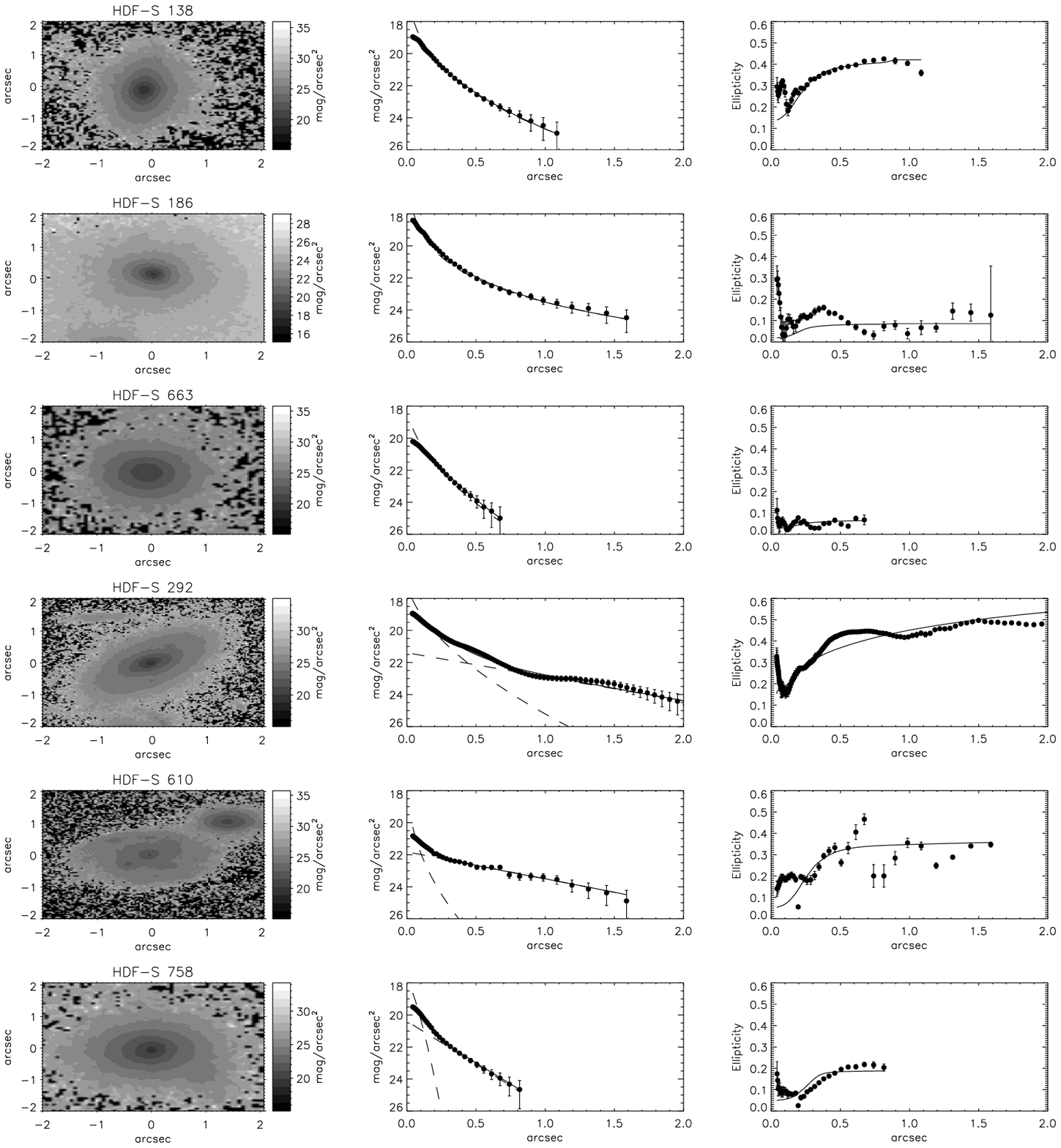}}

  \caption{Surface brightness and ellipticity semi--major radial profiles
  fitting to some galaxies in our sample. The three top galaxies are well
  fitted with a single S\'ersic law. The three bottom galaxies are fitted using
  a S\'ersic bulge plus an exponential disc.}

\label{fits}
  \end{figure*}

Our parameter--recovering method is described in detail in Trujillo et al.
(2001a) and Aguerri \& Trujillo (2002). The main points of the routine are as
follows. The final SB distributions resulting from the convolution between the
Point Spread Function (PSF) and our 2D S\'ersic model SB distribution are
dependent on the intrinsic ellipticity of the original source. However, to
evaluate the intrinsic ellipticity of a model (i.e. the value of the
ellipticity of the isophotes before being affected by the seeing) it is often
insufficient to simply measure the ellipticity at one given radial distance.
The ellipticity of the isophotes is reduced by seeing and this reduction
depends on the radial distance of the isophote to the centre of the model, the
size of the seeing, and the values of the model parameters (Trujillo et al.
2001b,c). For that reason, the determination of the intrinsic ellipticity of
the source and the fitting process to determine the structural parameters must
be done in tandem (i.e. iteratively and self--consistently). To do this we
simultaneously fit both the observed 1D SB and ellipticity semi--major axis
radial profiles using convolved profiles for each. We illustrate our technique
in Fig. \ref{fits}.

 The measured SB and ellipticity semi--major radial profiles are obtained from
fitting ellipses to the isophotes of the selected objects. The elliptical
isophotes are measured using the   task ELLIPSE within IRAF\footnote{IRAF is
distributed by the National Optical Astronomical Observatories, which are
operated by AURA, Inc. under contract to the NSF.} and fitting down to 1.5
$\sigma$ (where $\sigma$ is the standard deviation of the sky background of the
images).  A Levenberg--Marquardt non--linear fitting algorithm (Press et al.
1992) was used to determine the set of free parameters which minimizes
$\chi^2$. A variety of starting parameters are used to ensure that our fits do
not get trapped in local $\chi^2$ minima.

The PSF was estimated fitting a Moffat function to star profiles. The high
sampling of the  drizzled images that we used and the presence of enough
non--saturated stars along the field of view facilitates the estimation of the
PSF properties by a direct fit to this simple analytical function. We find that
a Moffat function with the parameters presented below reproduces  the observed
PSF very well. We have estimated a $\sim$5 per cent uncertainty in the estimation of
the FWHM due to changes from one WFPC2 position to another. In addition, the use
of an analytical PSF facilitates a robust analytical estimation of the
convolution between the PSF and the S\'ersic profile. 

 We find the following $\beta$ and FWHM (Full at Width Half Maximum) as our
best--fitting stellar parameters: $\beta$=3, FWHM=$0\farcs142$ (F606W);
$\beta$=2.5, FWHM=$0\farcs147$ (F814W). The best--fitting galaxy parameters in
the F814W band are shown in Table~\ref{data}. There, we list the following:

 \begin{enumerate}
 \item[(1)] The galaxy identification number as appearing in the photometric
redshift catalogs Fern\'andez--Soto, Lanzetta \& Yahil, (1999; HDF--N) and
Labb\'e et al. (2003; HDF--S).
\item[(2) and (3)] R.A. and Declination (J2000 coordinates).
\item[(4)] Best AB(8140) model--independent magnitudes according to the
photometric catalogs.
\item[(5)] The surface brightness of the bulge at the effective radius (mag/arcsec$^{2}$ in the I$_{814}$ band).
\item[(6)] The semi--major effective radius of the bulge (arcsec).
\item[(7)] The S\'ersic index $n$ of the bulge.
\item[(8)] The ellipticity of the bulge.
\item[(9)] The central surface brightness of the disc (mag/arcsec$^{2}$ in the I$_{814}$ band).
\item[(10)] The semi--major scale length of the disc (arcsec).
\item[(11)] The ellipticity of the disc.
\item[(12)] The bulge--to--total luminosity ratio ($B/T<0$ indicates galaxies 
classified as irregular/peculiar/merger systems).
\item[(13)] Spectroscopic redshift unless the index ``a'' appears which means
photometric redshift from Fern\'andez--Soto, Lanzetta \& Yahil, (1999; HDF--N)
and Labb\'e et al. (2003; HDF--S).
\item[(14)] Total AB(8140) magnitude as derived from the model fit.
\item[(15)] The global semi--major effective radius of the galaxy (arcsec).
 \end{enumerate}

\begin{table*}

 \centering
 \begin{minipage}{170mm}
  \caption{Best Morphological Parameter Values  For
  Galaxies with I$_{814}$(AB)$<$24.5 mag and z$<$1 in the HDF--N and HDF--S.}
  \begin{tabular}{cccccccccccccccc}
  \hline
 (1)  & (2) & (3) & (4)  & (5) & (6)  & (7) & (8)  & (9)
   & (10) & (11) & (12) & (13)   & (14) &    (15)    \\ 
  ID & R.A.(J2000) & D.(J2000) &I$_{814}$(cat) & $\mu_{e,b}$ & r$_{e,b}$ & n & $\epsilon_b$ &
  $\mu_d$(0) & h & $\epsilon_d$ & B/T & z  & I$_{814}$(mod) &
  r$_e$   \\
 & (h:m:s) & ($\circ$:m:s) & mag  & mag/''$^2$ & ('')  & &  &
  mag/''$^2$ & ('') &  &  &    & mag &
    ('')    \\   
 \hline
&  &  &  &  &   & & HDF--N &
  &  &  &  &    &  &
   &     \\
\hline
1      & 12:36:57.47 & 62:12:10.5  &   21.2  &   20.3 &  0.10  &     2.91  &  0.20&	 20.99 &   0.36 &     0.34 &   0.36 &	  0.665     &  21.1 &	0.38  \\
6      & 12:36:57.45 & 62:12:11.8  &   23.2  &   21.3 &  0.06  &     1.09  &  0.21&	 22.11 &   0.24 &     0.14 &   0.20 &	  0.561     &  23.1 &	0.31  \\
7      & 12:36:57.60 & 62:12:12.3  &   23.1  &   24.2 &  0.08  &     1.38  &  0.71&	 22.91 &   0.36 &     0.26 &   0.01 &	  0.920$^a$ &  23.4 &	0.59  \\
10     & 12:36:51.95 & 62:11:55.3  &   23.7  &   23.9 &  0.12  &     0.55  &  0.35&	 23.45 &   0.52 &     0.38 &   0.05 &	  0.080$^a$ &  23.3 &	0.82  \\
11     & 12:36:56.42 & 62:12:09.2  &   23.2  &   0.00 &  0.00  &     0.00  &  0.00&	 21.70 &   0.33 &     0.61 &   0.00 &	  0.321     &  23.1 &	0.50  \\
12     & 12:36:44.74 & 62:11:33.4  &   24.4  &   23.5 &  0.16  &     2.95  &  0.11&	 0.000 &   0.00 &     0.00 &   1.00 &	  0.880$^a$ &  24.4 &	0.16  \\
18     & 12:36:49.24 & 62:11:48.4  &   22.9  &   0.00 &  0.00  &     0.00  &  0.00&	 20.92 &   0.21 &     0.37 &   0.00 &	  0.961     &  22.8 &	0.37  \\
33     & 12:36:59.37 & 62:12:21.6  &   24.0  &   25.1 &  0.01  &     10.6  &  0.59&	 22.85 &   0.31 &     0.38 &   0.00 &	  0.472     &  23.9 &	0.51  \\
54     & 12:36:58.65 & 62:12:21.7  &   23.3  &   0.00 &  0.00  &     0.00  &  0.00&	 20.69 &   0.14 &     0.30 &   0.00 &	  0.681     &  23.3 &	0.22  \\
55     & 12:36:41.97 & 62:11:30.3  &   23.3  &   0.00 &  0.00  &     0.00  &  0.00&	 23.15 &   0.55 &     0.67 &   0.00 &	  0.200$^a$ &  23.8 &	0.69  \\
81     & 12:36:45.30 & 62:11:42.8  &   23.9  &   23.5 &  0.53  &     0.30  &  0.64&	 0.000 &   0.00 &     0.00 &  -1.00 &	  0.558     &  23.7 &	0.53  \\
85     & 12:36:41.61 & 62:11:31.8  &   19.7  &   21.8 &  0.34  &     0.46  &  0.40&	 20.72 &   0.81 &     0.70 &   0.14 &	  0.089     &  20.3 &	1.15  \\
91     & 12:36:42.28 & 62:11:34.7  &   23.8  &   22.8 &  0.15  &     0.27  &  0.64&	 23.26 &   0.34 &     0.26 &   0.14 &	  0.760$^a$ &  23.7 &	0.48  \\
122    & 12:36:43.80 & 62:11:42.8  &   20.8  &   23.6 &  0.91   &     6.32  &  0.25&	0.000 &   0.00 &     0.00 &   1.00 &	 0.764     &  20.4 &   0.91  \\
124    & 12:36:46.50 & 62:11:51.3  &   21.9  &   20.9 &  0.16   &     2.30  &  0.13&	0.000 &   0.00 &     0.00 &   1.00 &	 0.504     &  21.9 &   0.16  \\
125    & 12:37:00.56 & 62:12:34.6  &   21.3  &   22.0 &  0.35   &     2.73  &  0.07&	0.000 &   0.00 &     0.00 &   1.00 &	 0.562     &  21.2 &   0.35  \\
138    & 12:36:52.01 & 62:12:09.6  &   22.9  &   21.8 &  0.14   &     0.53  &  0.28&	23.42 &   0.61 &     0.46 &   0.31 &	 0.456     &  22.7 &   0.62  \\
144    & 12:36:57.21 & 62:12:25.8  &   22.3  &   22.5 &  0.09   &     0.84  &  0.66&	22.28 &   0.49 &     0.18 &   0.02 &	 0.561     &  22.0 &   0.80  \\
183    & 12:36:41.33 & 62:11:41.0  &   22.4  &   23.7 &  0.09   &     10.7  &  0.10&	21.24 &   0.26 &     0.39 &   0.15 &	 0.585     &  22.5 &   0.40  \\
191    & 12:36:45.33 & 62:11:54.4  &   23.0  &   23.4 &  0.30	&     4.64  &  0.21&   0.000 &   0.00 &     0.00 &   1.00 &	0.372	  &  22.8 &   0.30  \\
197    & 12:36:45.05 & 62:11:54.1  &   24.2  &   26.4 &  0.12	&     8.75  &  0.23&   23.21 &   0.30 &     0.07 &   0.03 &	0.920$^a$ &  23.8 &   0.49  \\
205    & 12:36:47.50 & 62:12:02.5  &   24.4  &   24.1 &  0.40	&     0.43  &  0.49&   0.000 &   0.00 &     0.00 &  -1.00 &	0.920$^a$ &  24.4 &   0.40  \\
218    & 12:36:46.26 & 62:11:59.7  &   24.4  &   0.00 &  0.00	&     0.00  &  0.00&   23.28 &   0.39 &     0.65 &   0.00 &	0.440$^a$ &  24.4 &   0.59  \\
220    & 12:36:52.68 & 62:12:19.6  &   23.2  &   21.6 &  0.20	&     0.90  &  0.33&   0.000 &   0.00 &     0.00 &   1.00 &	0.401	  &  22.9 &   0.20  \\
245    & 12:36:46.54 & 62:12:03.1  &   24.3  &   26.1 &  0.55	&     0.68  &  0.38&   20.74 &   0.09 &     0.50 &   0.32 &	0.454	  &  24.2 &   0.21  \\
247    & 12:36:45.95 & 62:12:01.3  &   23.7  &   23.0 &  0.41	&     0.47  &  0.54&   0.000 &   0.00 &     0.00 &  -1.00 &	0.679	  &  23.4 &   0.41  \\
257    & 12:36:51.72 & 62:12:20.1  &   21.5  &   0.00 &  0.00	&     0.00  &  0.00&   19.77 &   0.30 &     0.63 &   0.00 &	0.401	  &  21.4 &   0.46  \\
270    & 12:36:44.82 & 62:12:00.2  &   22.9  &   23.0 &  0.09	&     0.85  &  0.42&   22.67 &   0.42 &     0.24 &   0.05 &	0.457	  &  22.7 &   0.67  \\
273    & 12:36:50.16 & 62:12:16.9  &   22.4  &   0.00 &  0.00	&     0.00  &  0.00&   20.86 &   0.41 &     0.80 &   0.00 &	0.905	  &  22.5 &   0.58  \\
285    & 12:36:48.65 & 62:12:14.1  &   24.4  &   24.4 &  0.51	&     0.72  &  0.58&   0.000 &   0.00 &     0.00 &  -1.00 &	0.800$^a$ &  24.3 &   0.51  \\
310    & 12:36:48.12 & 62:12:14.8  &   24.4  &   22.2 &  0.13	&     0.77  &  0.35&   0.000 &   0.00 &     0.00 &   1.00 &	0.962	  &  24.5 &   0.13  \\
319    & 12:36:49.52 & 62:12:20.0  &   24.4  &   23.7 &  0.35	&     0.93  &  0.58&   0.000 &   0.00 &     0.00 &  -1.00 &	0.961	  &  24.2 &   0.35  \\
335    & 12:36:53.45 & 62:12:34.2  &   22.8  &   24.2 &  0.25	&     1.28  &  0.78&	  20.94 &   0.31 &     0.62 &	0.03 & 0.559	 &  22.4 &   0.51  \\
339    & 12:36:49.05 & 62:12:21.2  &   22.4  &   26.0 &  0.32	&     1.62  & 0.43&	  21.77 &   0.31 &     0.14 &	-0.03 & 0.953	 &  22.4 &   0.51  \\
344    & 12:36:50.85 & 62:12:27.2  &   24.2  &   24.2 &  0.39	&     0.49  &  0.33&	  0.000 &   0.00 &     0.00 &  -1.00 & 0.640$^a$ &  24.2 &   0.39  \\
345    & 12:36:56.65 & 62:12:45.3  &   20.0  &   21.2 &  0.27   &     1.06  &  0.10&	22.15 &   1.89 &     0.63 &   0.19 &	 0.517     &  19.6 &   2.49  \\
350    & 12:36:58.77 & 62:12:52.3  &   21.2  &   21.9 &  0.08   &     1.46  &  0.64&	21.36 &   0.48 &     0.32 &   0.02 &	 0.320     &  21.3 &   0.79  \\
351    & 12:36:58.32 & 62:12:51.1  &   24.0  &   24.9 &  0.15   &     1.55  &  0.12&	23.51 &   0.49 &     0.52 &   0.10 &	 0.240$^a$ &  23.7 &   0.74  \\
363    & 12:36:49.99 & 62:12:26.2  &   24.4  &   0.00 &  0.00	&     0.00  &  0.00&   21.74 &   0.23 &     0.78 &   0.00 &	0.010$^a$ &  24.5 &   0.38  \\
381    & 12:36:55.57 & 62:12:45.4  &   21.9  &   20.8 &  0.06	&     0.50  &  0.11&   21.31 &   0.31 &     0.22 &   0.09 &	0.790	  &  22.0 &   0.47  \\
391    & 12:36:41.94 & 62:12:05.4  &   21.0  &   0.00 &  0.00	&     0.00  &  0.00&   19.90 &   0.26 &     0.06 &   0.00 &	0.432	  &  20.8 &   0.46  \\
394    & 12:36:58.35 & 62:12:56.2  &   24.2  &   0.00 &  0.00	&     0.00  &  0.00&   22.25 &   0.19 &     0.49 &   0.00 &	0.520$^a$ &  24.5 &   0.31  \\
409    & 12:36:55.62 & 62:12:49.2  &   22.9  &   23.9 &  0.12	&     0.61  &  0.46&   23.56 &   0.62 &     0.31 &   0.03 &	0.950	  &  22.9 &   1.01  \\
424    & 12:36:40.95 & 62:12:05.3  &   23.2  &   22.0 &  0.14   &     2.48  &  0.14&	0.000 &   0.00 &     0.00 &   1.00 &	 0.882     &  23.3 &   0.13  \\
445    & 12:36:58.07 & 62:13:00.3  &   22.3  &   21.8 &  0.25   &     0.89  &  0.10&	23.78 &   0.95 &     0.60 &   0.63 &	 0.319     &  21.8 &   0.39  \\
450    & 12:36:55.26 & 62:12:52.5  &   24.2  &   22.9 &  0.01   &     10.2  &  0.99&	22.93 &   0.41 &     0.77 &   0.00 &	 0.720$^a$ &  24.4 &   0.68  \\
454    & 12:36:56.93 & 62:12:58.1  &   23.5  &   0.00 &  0.00   &     0.00  &  0.00&	21.25 &   0.21 &     0.54 &   0.00 &	 0.520     &  23.4 &   0.35  \\
458    & 12:36:57.31 & 62:12:59.6  &   21.3  &   23.9 &  0.48	&     2.37  &  0.88&   20.66 &   0.27 &     0.08 &   0.05 &	0.473	  &  21.5 &   0.45  \\
465    & 12:36:43.62 & 62:12:18.2  &   22.7  &   0.00 &  0.00	&     0.00  &  0.00&   20.70 &   0.19 &     0.31 &   0.00 &	0.750	  &  22.6 &   0.31  \\
467    & 12:36:42.91 & 62:12:16.3  &   20.7  &   21.8 &  0.52	&     0.45  &  0.36&   21.48 &   0.40 &     0.38 &   0.62 &	0.454	  &  20.9 &   0.56  \\
476    & 12:36:47.28 & 62:12:30.6  &   22.7  &   22.8 &  0.38	&     1.08  &  0.68&   23.51 &   0.68 &     0.25 &   0.32 &	0.421	  &  22.2 &   0.80  \\
477    & 12:36:50.21 & 62:12:39.7  &   20.6  &   25.0 &  0.20	&     10.0  &  0.14&   20.88 &   0.60 &     0.26 &  -0.03 &	0.474	  &  20.2 &   0.99  \\
495    & 12:36:41.48 & 62:12:14.9  &   23.4  &   20.1 &  0.06	&     1.44  &  0.69&   20.66 &   0.13 &     0.43 &   0.28 &	0.880$^a$ &  23.3 &   0.15  \\
500    & 12:36:53.90 & 62:12:53.9  &   20.8  &   20.5 &  0.09	&     1.15  &  0.58&   20.83 &   0.43 &     0.27 &   0.06 &	0.642	  &  20.9 &   0.68  \\
512    & 12:36:54.78 & 62:12:58.2  &   24.3  &   24.1 &  0.34	&     0.51  &  0.32&   0.000 &   0.00 &     0.00 &  -1.00 &	0.851	  &  24.4 &   0.34  \\
521    & 12:36:39.43 & 62:12:11.7  &   23.8  &   23.0 &  0.20	&     3.41  &  0.25&   0.000 &   0.00 &     0.00 &   1.00 &	1.000$^a$ &  23.5 &   0.19  \\
524    & 12:36:50.26 & 62:12:45.7  &   21.4  &   22.2 &  0.23	&     6.31  &  0.22&   21.46 &   0.38 &     0.38 &   0.50 &	0.678	  &  21.3 &   0.46  \\
537    & 12:36:47.04 & 62:12:36.8  &   20.9  &   20.8 &  0.19	&     0.98  &  0.36&   21.36 &   0.40 &     0.09 &   0.32 &	0.321	  &  21.0 &   0.44  \\
549    & 12:36:55.16 & 62:13:03.6  &   23.5  &   21.7 &  0.14	&     2.05  &  0.29&   0.000 &   0.00 &     0.00 &   1.00 &	0.952	  &  23.3 &   0.13  \\
563    & 12:36:50.84 & 62:12:51.4  &   23.1  &   26.5 &  0.16	&     5.17  &  0.28&   23.54 &   0.80 &     0.26 &   0.01 &	0.485	  &  22.3 &   1.33  \\
\hline
\label{data}
\end{tabular}

\end{minipage}

\end{table*}

\begin{table*}

 \centering
 \begin{minipage}{170mm}
  \begin{tabular}{cccccccccccccccc}
  \hline
   (1)  & (2) & (3) & (4)  & (5) & (6)  & (7) & (8)  & (9)
   & (10) & (11) & (12) & (13)   & (14) &    (15)    \\ 
  ID & R.A.(J2000) & D.(J2000) &I$_{814}$(cat) & $\mu_{e,b}$ & r$_{e,b}$ & n & $\epsilon_b$ &
  $\mu_d$(0) & h & $\epsilon_d$ & B/T & z  & I$_{814}$(mod) &
  r$_e$   \\
 & (h:m:s) & ($\circ$:m:s) & mag  & mag/''$^2$ & ('')  & &  &
  mag/''$^2$ & ('') &  &  &    & mag &
    ('')    \\   
 \hline
565 & 12:36:48.99&62:12:45.8  &   23.6  &   23.4 &  0.41   &	 0.66  &  0.48&      0.000 &   0.00 &	  0.00 &  -1.00 &     0.512   &  23.5 &   0.41  \\
581 & 12:36:49.47&62:12:48.7  &   23.9  &   22.6 &  0.12   &	 0.61  &  0.03&      22.91 &   0.21 &	  0.16 &   0.45 &     0.760$^a$ &  23.8 &   0.19  \\
599 & 12:36:57.71&62:13:15.1  &   23.0  &   24.1 &  0.51   &	 1.16  &  0.05&       0.000 &	0.00 &     0.00 &  -1.00 &     0.952   &  22.8 &   0.51  \\
611 & 12:36:50.82&62:12:55.8  &   22.5  &   22.3 &  0.12   &	 2.03  &  0.37&      20.92 &   0.30 &	  0.49 &   0.13 &     0.319   &  22.1 &   0.44  \\
617 & 12:36:38.97&62:12:19.7  &   22.2  &   23.2 &  0.48   &	 0.98  &  0.05&      0.000 &   0.00 &	  0.00 &  -1.00 &     0.609   &  22.2 &   0.48  \\
619 & 12:36:55.46&62:13:11.1  &   22.1  &   24.3 &  0.72   &	 3.98  &  0.05&       0.000 &	0.00 &     0.00 &   1.00 &     0.370   &  21.7 &   0.72  \\
631 & 12:36:55.15&62:13:11.3  &   23.6  &   22.4 &  0.23   &	 0.51  &  0.17&       0.000 &	0.00 &     0.00 &  -1.00 &     0.321   &  23.4 &   0.23  \\
637 & 12:36:51.43&62:13:00.6  &   22.7  &   25.0 &  0.89   &	 0.70  &  0.23&      0.000 &   0.00 &	  0.00 &  -1.00 &     0.089   &  22.9 &   0.89  \\
650 & 12:36:44.18&62:12:40.3  &   23.3  &   23.6 &  0.08   &	 1.08  &  0.11&      23.12 &   0.47 &	  0.05 &   0.03 &     0.873   &  22.7 &   0.76  \\
653 & 12:36:46.12&62:12:46.4  &   22.3  &   21.0 &  0.10   &	 2.20  &  0.08&      22.73 &   0.43 &	  0.25 &   0.45 &     0.900   &  22.2 &   0.33  \\
655 & 12:36:49.63&62:12:57.5  &   21.9  &   0.00 &  0.00   &	 0.00  &  0.00&      20.06 &   0.26 &	  0.59 &   0.00 &     0.477   &  21.9 &   0.42  \\
662 & 12:36:55.00&62:13:14.7  &   23.8  &   23.7 &  0.22   &	 0.37  &  0.23&       22.02 &	0.22 &     0.63 &   0.35 &     0.511   &  23.9 &   0.29  \\
669 & 12:36:39.77&62:12:28.5  &   23.7  &   0.00 &  0.00   &	 0.00  &  0.00&      22.31 &   0.26 &	  0.25 &   0.00 &     0.010$^a$ &  23.6 &   0.37  \\
671 & 12:36:47.54&62:12:52.6  &   23.9  &   0.00 &  0.00   &	 0.00  &  0.00&      22.07 &   0.32 &	  0.66 &   0.00 &     0.681   &  23.6 &   0.55  \\
683 & 12:36:41.61&62:12:35.6  &   24.4  &   22.6 &  0.13   &	 1.76  &  0.24&      0.000 &   0.00 &	  0.00 &   1.00 &     0.520$^a$ &  24.3 &   0.12  \\
689 & 12:36:39.86&62:12:31.5  &   24.1  &   21.9 &  0.11   &	 0.66  &  0.60&      23.07 &   0.32 &	  0.52 &   0.30 &     0.640$^a$ &  23.9 &   0.34  \\
694 & 12:36:43.15&62:12:42.2  &   21.7  &   23.2 &  0.57   &	 5.70  &  0.25&      0.000 &   0.00 &	  0.00 &   1.00 &     0.847   &  21.2 &   0.56  \\
709 & 12:36:44.18&62:12:47.8  &   21.6  &   21.7 &  0.10   &	 1.79  &  0.20&      21.25 &   0.36 &	  0.20 &   0.11 &     0.558   &  21.5 &   0.53  \\
716 & 12:36:38.40&62:12:31.3  &   22.6  &   0.00 &  0.00   &	 0.00  &  0.00&      21.87 &   0.58 &	  0.77 &   0.00 &     0.944   &  22.6 &   0.90  \\
719 & 12:36:49.68&62:13:07.3  &   24.2  &   24.2 &  0.18   &	 0.85  &  0.32&      22.51 &   0.17 &	  0.18 &   0.25 &     0.920$^a$ &  24.2 &   0.25  \\
720 & 12:36:43.96&62:12:50.0  &   20.9  &   23.7 &  0.37   &	 1.79  &  0.60&      20.58 &   0.53 &	  0.60 &   -0.06 &     0.557   &  20.8 &   0.85  \\
727 & 12:36:38.60&62:12:33.8  &   23.9  &   24.4 &  0.06   &	 2.85  &  0.61&      21.95 &   0.30 &	  0.66 &   0.02 &     0.904   &  23.7 &   0.49  \\
740 & 12:36:49.03&62:13:09.7  &   24.0  &   21.9 &  0.18   &	 0.46  &  0.44&      0.000 &   0.00 &	  0.00 &   1.00 &     0.960$^a$ &  23.8 &   0.18  \\
743 & 12:36:49.37&62:13:11.2  &   22.1  &   21.3 &  0.20   &	 2.50  &  0.15&      0.000 &   0.00 &	  0.00 &   1.00 &     0.478   &  21.8 &   0.19  \\
746 & 12:36:49.72&62:13:13.0  &   21.4  &   24.5 &  0.77   &	 1.34  &  0.70&      21.09 &   0.57 &	  0.70 &   0.13 &     0.475   &  21.4 &   0.93  \\
749 & 12:36:50.47&62:13:16.1  &   22.6  &   22.6 &  0.04   &	 17.0  &  0.62&      22.39 &   0.67 &	  0.71 &   0.03 &     0.851   &  22.6 &   1.09  \\
751 & 12:36:48.08&62:13:09.0  &   20.4  &   19.1 &  0.10   &	 1.52  &  0.04&      21.58 &   0.57 &	  0.08 &   0.41 &     0.475   &  20.2 &   0.42  \\
757 & 12:36:51.05&62:13:20.6  &   19.6  &   21.7 &  0.28   &	 0.58  &  0.60&      20.99 &   1.29 &	  0.60 &   0.03 &     0.199   &  19.3 &   2.10  \\
774 & 12:36:44.58&62:13:04.6  &   21.2  &   19.7 &  0.05   &	 2.32  &  0.72&      20.53 &   0.53 &	  0.71 &   0.05 &     0.485   &  21.1 &   0.84  \\
775 & 12:36:48.47&62:13:16.6  &   23.2  &   22.2 &  0.17   &	 0.99  &  0.48&      23.15 &   0.32 &	  0.09 &   0.41 &     0.474   &  23.1 &   0.32  \\
780 & 12:36:48.78&62:13:18.4  &   22.6  &   23.2 &  0.07   &	 0.62  &  0.14&      23.12 &   0.64 &	  0.25 &   0.02 &     0.749   &  22.3 &   1.05  \\
781 & 12:36:46.75&62:13:12.3  &   23.6  &   23.2 &  0.09   &	 0.77  &  0.40&      22.32 &   0.38 &	  0.63 &   0.07 &     0.600$^a$ &  23.4 &   0.59  \\
782 & 12:36:45.61&62:13:08.9  &   23.6  &   22.6 &  0.18   &	 2.20  &  0.18&      0.000 &   0.00 &	  0.00 &   1.00 &     0.480$^a$ &  23.4 &   0.17  \\
793 & 12:36:49.07&62:13:21.8  &   24.4  &   23.5 &  0.30   &	 0.40  &  0.05&      0.000 &   0.00 &	  0.00 &  -1.00 &     0.880$^a$ &  23.8 &   0.30  \\
798 & 12:36:46.15&62:13:13.8  &   24.2  &   21.6 &  0.11   &	 1.14  &  0.31&      0.000 &   0.00 &	  0.00 &   1.00 &     0.640$^a$ &  24.1 &   0.11  \\
810 & 12:36:42.51&62:13:05.1  &   24.3  &   24.5 &  0.25   &	 3.00  &  0.20&      0.000 &   0.00 &	  0.00 &  -1.00 &     0.720$^a$ &  24.5 &   0.25  \\
817 & 12:36:42.72&62:13:07.1  &   22.1  &   28.4 &  1.88   &	 10.0  &  0.67&      19.82 &   0.24 &	  0.74 &   0.20 &     0.485   &  22.2 &   0.44  \\
866 & 12:36:47.45&62:13:30.0  &   23.5  &   0.00 &  0.00   &	 0.00  &  0.00&      22.57 &   0.26 &	  0.02 &   0.00 &     0.960$^a$ &  23.5 &   0.37  \\
869 & 12:36:45.86&62:13:25.8  &   20.9  &   24.8 &  0.32   &	 1.55  &  0.51&      21.96 &   0.84 &	  0.28 &   0.01 &     0.321   &  20.6 &   1.39  \\
876 & 12:36:45.41&62:13:25.9  &   22.6  &   20.1 &  0.10   &	 3.04  &  0.67&      20.34 &   0.18 &	  0.52 &   0.43 &     0.441   &  22.2 &   0.21  \\
882 & 12:36:52.23&62:13:48.0  &   24.3  &   22.8 &  0.10   &	 0.68  &  0.23&       0.000 &	0.00 &     0.00 &   1.00 &     0.520$^a$ &  24.2 &   0.19  \\
884 & 12:36:54.10&62:13:54.3  &   22.4  &   0.00 &  0.00   &	 0.00  &  0.00&       21.46 &	0.32 &     0.29 &   0.00 &     0.849   &  22.3 &   0.54  \\
888 & 12:36:55.59&62:13:59.8  &   23.7  &   22.3 &  0.03   &	 6.45  &  0.10&       22.41 &	0.40 &     0.72 &   0.07 &     0.559   &  23.6 &   0.62  \\
890 & 12:36:42.37&62:13:19.3  &   23.8  &   21.6 &  0.10   &	 2.51  &  0.24&       0.000 &	0.00 & 0.00 &	1.00 &     0.760$^a$ &  23.8 &   0.09  \\
899 & 12:36:55.50&62:14:02.6  &   22.9  &   0.00 &  0.00   &	 0.00  &  0.00&       21.48 &	0.47 &     0.80 &   0.00 &     0.564   &  22.8 &   0.80  \\
902 & 12:36:42.02&62:13:21.4  &   23.5  &   21.0 &  0.09   &	 1.25  &  0.64&       22.77 &	0.20 &     0.05 &   0.46 &     0.846   &  23.6 &   0.18  \\
912 & 12:36:51.80&62:13:53.8  &   21.1  &   21.8 &  0.56   &	 0.58  &  0.60&       22.25 &	0.86 &     0.60 &  -0.49 &     0.557   &  20.8 &   0.82  \\
914 & 12:36:49.44&62:13:46.9  &   18.1  &   19.4 &  0.56   &	 2.40  &  0.40&      0.000 &   0.00 &	  0.00 &   1.00 &     0.089   &  18.1 &   0.55  \\
915 & 12:36:48.03&62:13:43.0  &   24.3  &   24.8 &  0.15   &	 0.45  &  0.60&      23.80 &   0.55 &	  0.60 &   0.04 &     0.840$^a$ &  24.0 &   0.88  \\
938 & 12:36:51.98&62:14:00.8  &   23.0  &   0.00 &  0.00   &	 0.00  &  0.00&       21.02 &	0.32 &     0.67 &   0.00 &     0.557   &  22.7 &   0.53  \\
941 & 12:36:52.87&62:14:04.8  &   23.2  &   25.1 &  0.19   &	 1.52  &  0.14&       21.96 &	0.21 &     0.05 &  -0.08 &     0.498   &  23.3 &   0.33  \\
979 & 12:36:53.65&62:14:17.6  &   23.4  &   0.00 &  0.00   &	 0.00  &  0.00&       21.36 &	0.33 &     0.73 &   0.00 &     0.517   &  23.2 &   0.48  \\
989 & 12:36:49.50&62:14:06.7  &   21.8  &   21.1 &  0.18   &	 0.61  &  0.48&      21.29 &   0.36 &	  0.31 &   0.25 &     0.752   &  21.6 &   0.44  \\
990 & 12:36:46.03&62:13:56.3  &   24.2  &   0.00 &  0.00   &	 0.00  &  0.00&      21.93 &   0.16 &	  0.31 &   0.00 &     1.000$^a$ &  24.2 &   0.25  \\
1018& 12:36:46.35&62:14:04.6  &   21.2  &   24.8 &  1.13   &	 11.0  &  0.05&      0.000 &   0.00 &	  0.00 &   1.00 &     0.960   &  20.6 &   1.12  \\
1022& 12:36:51.39&62:14:20.9  &   23.3  &   0.00 &  0.00   &	 0.00  &  0.00&      22.05 &   0.53 &	  0.60 &   0.00 &     0.439   &  22.8 &   0.88  \\
1027& 12:36:48.35&62:14:12.3  &   23.8  &   24.2 &  0.88   &	 0.43  &  0.80&       0.000 &	0.00 &     0.00 &  -1.00 &     0.920$^a$ &  23.9 &   0.88  \\
1028& 12:36:50.36&62:14:18.6  &   23.0  &   23.1 &  0.18   &	 0.79  &  0.84&       22.45 &	0.39 &     0.37 &   0.05 &     0.816   &  22.9 &   0.62  \\
1029& 12:36:46.54&62:14:07.5  &   23.9  &   22.7 &  0.20   &	 1.54  &  0.33&      0.000 &   0.00 &	 0.00 &   1.00 &     0.130   &  23.7 &   0.19  \\
1038& 12:36:48.10&62:14:14.4  &   24.4  &   21.8 &  0.06   &	 0.57  &  0.24&      22.03 &   0.15 &	 0.37 &   0.29 &     0.920$^a$ &  24.3 &   0.16  \\
1048& 12:36:47.18&62:14:14.2  &   23.5  &   23.1 &  0.15   &	 0.53  &  0.49&      21.06 &   0.18 &	 0.43 &   0.11 &     0.609   &  23.2 &   0.27  \\
1063& 12:36:53.11&62:14:38.0  &   24.2  &   23.5 &  0.37   &	 0.85  &  0.60&      0.000 &   0.00 &	  0.00 &  -1.00 &     0.400$^a$ &  24.0 &   0.37  \\
1067& 12:36:48.32&62:14:26.2  &   19.1  &   19.8 &  0.18   &	 1.02  &  0.20&      19.40 &   0.48 &	  0.20 &   0.15 &     0.139   &  19.0 &   0.68  \\
\hline

\end{tabular}

\end{minipage}

\end{table*}

\begin{table*}

 \centering
 \begin{minipage}{170mm}
  \begin{tabular}{cccccccccccccccc}
  \hline
   (1)  & (2) & (3) & (4)  & (5) & (6)  & (7) & (8)  & (9)
   & (10) & (11) & (12) & (13)   & (14) &    (15)    \\ 
  ID & R.A.(J2000) & D.(J2000) &I$_{814}$(cat) & $\mu_{e,b}$ & r$_{e,b}$ & n & $\epsilon_b$ &
  $\mu_d$(0) & h & $\epsilon_d$ & B/T & z  & I$_{814}$(mod) &
  r$_e$   \\
 & (h:m:s) & ($\circ$:m:s) & mag  & mag/''$^2$ & ('')  & &  &
  mag/''$^2$ & ('') &  &  &    & mag &
    ('')    \\   
 
 \hline
&  &  &  &  &   & & HDF--S &
  &  &  &  &    &  &
   &     \\
\hline

29  & 22:32:46.27&-60:34:16.3  & 21.0 &  19.3 &    0.09  &  0.49  &  0.23 &  20.19 &  0.19 &  0.03  &	-0.36	&     0.460$^a$  &   21.2  &   0.19  \\
30  & 22:33:00.48&-60:34:17.6  & 23.2 &  24.3 &    0.45  &  4.82  &  0.29 &  0.000 &  0.00 &  0.00  &	 1.00	&     0.620$^a$  &   22.9  &   0.45  \\
41  & 22:32:56.06&-60:34:14.1  & 22.1 &  21.4 &    0.22  &  0.50  &  0.47 &  22.32 &  0.45 &  0.33  &	 0.39	&     0.564	 &   22.0  &   0.42  \\
49  & 22:32:45.60&-60:34:15.1  & 24.4 &  24.8 &    0.50  &  2.04  &  0.41 &  0.000 &  0.00 &  0.00  &	-1.00	&     0.462	 &   23.8  &   0.50  \\
50  & 22:32:47.55&-60:34:08.5  & 21.2 &  0.00 &    0.00  &  0.00  &  0.00 &  20.53 &  0.34 &  0.23  &	 0.00	&     0.560	 &   21.2  &   0.60  \\
53  & 22:32:50.94&-60:34:15.2  & 24.2 &  0.00 &    0.00  &  0.00  &  0.00 &  20.51 &  0.09 &  0.19  &	 0.00	&     0.424	 &   23.8  &   0.15  \\
64  & 22:32:55.21&-60:34:10.0  & 23.4 &  0.00 &    0.00  &  0.00  &  0.00 &  21.99 &  0.22 &  0.05  &	 0.00	&     0.960$^a$  &   23.3  &   0.32  \\
65  & 22:33:01.54&-60:34:10.7  & 23.5 &  24.7 &    0.13  &  0.79  &  0.63 &  21.50 &  0.20 &  0.41  &	 0.02	&     0.853	  &   23.5  &	0.33  \\
75  & 22:32:55.23&-60:34:07.4  & 23.1 &  23.9 &    0.10  &  0.83  &  0.35 &  22.69 &  0.38 &  0.20  &	 0.03	&     0.465	  &   22.9  &	0.62  \\
87  & 22:33:04.44&-60:34:09.0  & 24.0 &  24.0 &    0.18  &  0.69  &  0.10 &  23.13 &  0.31 &  0.29  &	 0.22	&     0.480$^a$   &   23.7  &	0.41  \\
91  & 22:32:52.22&-60:34:02.6  & 22.0 &  21.8 &    0.11  &  0.72  &  0.27 &  20.44 &  0.26 &  0.63  &	 0.13	&     0.511	 &   22.3  &   0.37  \\
95  & 22:32:45.98&-60:34:07.2  & 23.8 &  0.00 &    0.00  &  0.00  &  0.00 &  19.53 &  0.07 &  0.25  &	 0.00	&     0.753	 &   23.6  &   0.12  \\
97  & 22:32:48.89&-60:34:04.6  & 22.6 &  20.7 &    0.09  &  3.36  &  0.03 &  0.000 &  0.00 &  0.00  &	 1.00	&     0.980$^a$  &   22.5  &   0.09  \\
99  & 22:32:52.13&-60:33:59.3  & 21.8 &  23.5 &    0.23  &  1.47  &  0.50 &  21.94 &  0.86 &  0.59  &	 0.04	&     0.564	 &   21.1  &   1.38  \\
138 & 22:32:46.88&-60:33:54.7  & 21.2 &  20.2 &    0.19  &  3.07  &  0.42 &  0.000 &  0.00 &  0.00  &	 1.00	&     0.420$^a$  &   21.1  & 0.19  \\
153 & 22:32:48.23&-60:33:54.9  & 23.4 &  0.00 &    0.00  &  0.00  &  0.00 &  21.93 &  0.39 &  0.64  &	 0.00	&     0.721    &   23.1  &   0.54  \\
184 & 22:33:03.56&-60:33:41.6  & 20.0 &  23.6 &    0.37  &  3.49  &  0.28 &  19.86 &  0.43 &  0.25  &	-0.07	&    0.340	&   19.8  &   0.71  \\
186 & 22:33:05.77&-60:33:41.5  & 20.0 &  22.8 &    0.72  &  7.84  &  0.05 &  0.000 &  0.00 &  0.00  &	 1.00	&    0.328	&   19.8  &   0.71  \\
189 & 22:32:57.87&-60:33:49.1  & 23.6 &  23.2 &    0.17  &  0.92  &  0.83 &  23.37 &  0.71 &  0.60  &	 0.05	&     0.428    &   23.0  &   1.13  \\
193 & 22:32:47.64&-60:33:35.7  & 19.5 &  20.7 &    0.19  &  3.08  &  0.18 &  20.20 &  0.58 &  0.27  &	 0.19	&     0.580    &   19.4  &   0.82  \\
198 & 22:33:02.44&-60:33:46.5  & 22.3 &  22.2 &    0.14  &  0.48  &  0.27 &  22.51 &  0.47 &  0.37  &	-0.16	&      0.695	  &   22.4  &	0.66  \\
215 & 22:32:47.47&-60:33:43.9  & 24.2 &  0.00 &    0.00  &  0.00  &  0.00 &  21.68 &  0.13 &  0.07  &	 0.00	&     0.560$^a$  &   24.1  & 0.24  \\
220 & 22:32:49.42&-60:33:43.6  & 24.2 &  0.00 &    0.00  &  0.00  &  0.00 &  21.04 &  0.10 &  0.25  &	 0.00	&     0.647    &   24.3  &   0.16  \\
223 & 22:32:53.73&-60:33:37.4  & 21.5 &  24.0 &    0.35  &  0.99  &  0.10 &  22.83 &  0.88 &  0.15  &	 0.09	&     0.565    &   21.1  &   1.34  \\
227 & 22:32:59.42&-60:33:39.6  & 22.1 &  0.00 &    0.00  &  0.00  &  0.00 &  21.88 &  0.43 &  0.28  &	 0.00	&     0.464    &   22.0  &   0.73  \\
234 & 22:32:51.50&-60:33:37.4  & 22.0 &  22.0 &    0.07  &  1.48  &  0.92 &  19.77 &  0.18 &  0.45  &	-0.01	&     0.577    &   22.1  &   0.30  \\
244 & 22:32:54.67&-60:33:33.0  & 21.2 &  22.1 &    0.02  &  1.29  &  0.42 &  20.15 &  0.52 &  0.75  &	 0.01	&     0.173    &   21.0  &   0.87  \\
247 & 22:32:45.30&-60:33:32.4  & 21.5 &  20.8 &    0.19  &  2.92  &  0.26 &  0.000 &  0.00 &  0.00  &	 1.00	&     0.520$^a$  &   21.5  & 0.19  \\
256 & 22:32:58.21&-60:33:31.5  & 22.0 &  0.00 &    0.00  &  0.00  &  0.00 &  21.51 &  0.37 &  0.25  &	 0.00	&     0.423    &   21.9  &   0.60  \\
265 & 22:32:55.72&-60:33:33.6  & 23.8 &  24.5 &    0.02  &  12.0  &  0.10 &  23.81 &  0.54 &  0.23  &	 0.00	&     0.564    &   23.4  &   0.54  \\
270 & 22:32:52.34&-60:33:32.9  & 22.7 &  0.00 &    0.00  &  0.00  &  0.00 &  19.48 &  0.11 &  0.23  &	 0.00	&     0.500$^a$  &   22.5  & 0.17  \\
273 & 22:33:02.12&-60:33:34.9  & 24.1 &  23.4 &    0.08  &  1.51  &  0.10 &  23.18 &  0.57 &  0.65  &	 0.08	&     0.500$^a$  &   23.4  & 0.89  \\
283 & 22:32:55.81&-60:33:30.2  & 24.0 &  0.00 &    0.00  &  0.00  &  0.00 &  22.92 &  0.51 &  0.62  &	 0.00	&     0.180$^a$  &   23.4  & 0.89  \\
288 & 22:33:04.25&-60:33:31.8  & 24.0 &  0.00 &    0.00  &  0.00  &  0.00 &  22.62 &  0.25 &  0.05  &	 0.00	&    0.465	&   23.6  &   0.40  \\
292 & 22:33:02.75&-60:33:22.0  & 20.2 &  20.1 &    0.20  &  1.75  &  0.36 &  21.40 &  0.72 &  0.57  &	 0.48	&    0.464	&   20.3  &   0.53  \\
297 & 22:32:57.08&-60:33:28.8  & 23.9 &  22.5 &    0.12  &  1.60  &  0.29 &  24.44 &  0.47 &  0.01  &	 0.39	&     0.680$^a$  &   23.5  & 0.41  \\
301 & 22:32:59.45&-60:33:28.8  & 24.2 &  22.4 &    0.13  &  1.51  &  0.64 &  21.63 &  0.16 &  0.51  &	 0.35	&     0.980$^a$  &   23.8  & 0.22  \\
306 & 22:33:00.86&-60:33:25.6  & 23.6 &  24.5 &    0.48  &  0.92  &  0.41 &  0.000 &  0.00 &  0.00  &	-1.00	&     0.940$^a$  &   23.9  & 0.48  \\
307 & 22:32:50.55&-60:33:25.8  & 22.6 &  26.4 &    0.13  &  2.71  &  0.10 &  21.36 &  0.29 &  0.29  &	 0.01	&     0.440$^a$  &   22.4  & 0.48  \\
314 & 22:32:52.15&-60:33:23.8  & 22.8 &  0.00 &    0.00  &  0.00  &  0.00 &  20.79 &  0.16 &  0.14  &	 0.00	&     0.220$^a$  &   22.8  & 0.25  \\
317 & 22:32:57.04&-60:33:22.9  & 22.8 &  20.8 &    0.09  &  2.61  &  0.16 &  0.000 &  0.00 &  0.00  &	 1.00	&     0.520$^a$  &   22.9  & 0.09  \\
337 & 22:32:45.77&-60:32:50.4  & 22.1 &  0.00 &    0.00  &  0.00  &  0.00 &  21.81 &  0.33 &  0.06  &	 0.00	&     0.852    &   22.3  &   0.53  \\
339 & 22:32:54.04&-60:32:51.6  & 22.1 &  20.7 &    0.14  &  1.51  &  0.10 &  21.70 &  0.32 &  0.37  &	 0.62	&     0.515    &   21.5  &   0.23  \\
344 & 22:32:47.66&-60:32:52.2  & 23.3 &  25.3 &    0.07  &  1.14  &  0.90 &  22.18 &  0.38 &  0.52  &	 0.01	&     0.480$^a$  &   23.0  & 0.64  \\
345 & 22:32:49.23&-60:32:53.3  & 22.5 &  24.3 &    0.22  &  1.98  &  0.11 &  22.76 &  0.58 &  0.03  &	 0.08	&     0.500$^a$  &   21.8  & 0.90  \\
367 & 22:32:47.77&-60:32:55.9  & 24.2 &  26.7 &    0.07  &  10.5  &  0.10 &  23.96 &  0.42 &  0.21  &	-0.04	&     0.960$^a$  &   24.0  & 0.07  \\
374 & 22:33:00.48&-60:32:58.4  & 24.1 &  0.00 &    0.00  &  0.00  &  0.00 &  20.68 &  0.10 &  0.20  &	 0.00	&     0.440$^a$  &   23.9  & 0.14  \\
394 & 22:32:45.63&-60:33:01.3  & 21.4 &  0.00 &    0.00  &  0.00  &  0.00 &  19.81 &  0.24 &  0.40  &	 0.00	&     0.760$^a$  &   21.4  & 0.41  \\
409 & 22:32:57.53&-60:33:06.0  & 20.4 &  21.9 &    0.05  &  4.38  &  0.01 &  20.71 &  0.40 &  0.01  &	-0.02	&     0.580    &   20.6  &   0.66  \\
410 & 22:32:57.21&-60:33:05.4  & 22.8 &  0.00 &    0.00  &  0.00  &  0.00 &  22.00 &  0.40 &  0.46  &	 0.00	&     0.582    &   22.7  &   0.56  \\
415 & 22:32:54.01&-60:33:05.4  & 22.7 &  0.00 &    0.00  &  0.00  &  0.00 &  20.42 &  0.29 &  0.72  &	 0.00	&     0.581    &   22.4  &   0.49  \\
420 & 22:32:51.66&-60:33:05.9  & 24.3 &  23.2 &    0.16  &  3.31  &  0.14 &  0.000 &  0.00 &  0.00  &	 1.00	&     0.980$^a$  &   24.1  & 0.16  \\
427 & 22:32:52.29&-60:33:08.3  & 21.1 &  21.4 &    0.09  &  0.63  &  0.61 &  21.74 &  0.64 &  0.53  &	 0.03	&     0.583    &   21.5  &   1.05  \\
429 & 22:33:03.85&-60:33:08.7  & 22.9 &  25.1 &    0.16  &  12.4  &  0.10 &  22.69 &  0.45 &  0.46  &	 0.13	&    0.970	&   22.9  &   0.71  \\
438 & 22:32:48.99&-60:33:09.2  & 24.6 &  0.00 &    0.00  &  0.00  &  0.00 &  22.17 &  0.22 &  0.57  &	 0.00	&     0.359    &   24.3  &   0.36  \\
442 & 22:32:49.49&-60:33:11.0  & 23.1 &  23.9 &    0.33  &  1.24  &  0.14 &  23.94 &  0.72 &  0.33  &	 0.36	&     0.564    &   22.5  &   0.78  \\
446 & 22:32:53.91&-60:33:13.3  & 21.0 &  22.6 &    0.26  &  4.05  &  0.06 &  20.10 &  0.39 &  0.40  &	 0.33	&     0.364    &   20.4  &   0.54  \\
456 & 22:32:50.97&-60:33:13.9  & 24.4 &  22.2 &    0.04  &  3.75  &  0.10 &  22.67 &  0.35 &  0.61  &	 0.15	&     0.460$^a$  &   23.7  & 0.51  \\
459 & 22:32:48.55&-60:33:13.8  & 23.3 &  0.00 &    0.00  &  0.00  &  0.00 &  19.20 &  0.07 &  0.29  &	 0.00	&     0.480$^a$  &   23.1  & 0.14  \\
464 & 22:33:01.88&-60:33:16.3  & 21.7 &  24.1 &    0.57  &  2.32  &  0.88 &  20.13 &  0.26 &  0.52  &	 0.07	&    0.428     &   21.7  &   0.45  \\
476 & 22:32:55.86&-60:33:17.6  & 23.0 &  25.6 &    0.11  &  12.7  &  0.47 &  21.35 &  0.20 &  0.19  &	 0.04	&    0.581     &   23.0  &   0.33  \\
482 & 22:33:00.14&-60:33:18.9  & 23.3 &  21.8 &    0.11  &  0.83  &  0.28 &  22.68 &  0.28 &  0.02  &	 0.31	&    0.540     &   23.0  &   0.30  \\
\hline

\end{tabular}

\end{minipage}

\end{table*}

\begin{table*}

 \centering
 \begin{minipage}{170mm}
  \begin{tabular}{cccccccccccccccc}
  \hline
   (1)  & (2) & (3) & (4)  & (5) & (6)  & (7) & (8)  & (9)
   & (10) & (11) & (12) & (13)   & (14) &    (15)    \\ 
  ID & R.A.(J2000) & D.(J2000) &I$_{814}$(cat) & $\mu_{e,b}$ & r$_{e,b}$ & n & $\epsilon_b$ &
  $\mu_d$(0) & h & $\epsilon_d$ & B/T & z  & I$_{814}$(mod) &
  r$_e$  \\
 & (h:m:s) & ($\circ$:m:s) & mag  & mag/''$^2$ & ('')  & &  &
  mag/''$^2$ & ('') &  &  &    & mag &
    ('')    \\   
 \hline
487   &   22:32:52.25& -60:33:19.3  & 24.0 &  23.5 &	0.34  &  0.33  &  0.19 &  0.000 &  0.00 &  0.00  & -1.00   &	0.240$^a$	&   23.7  &   0.34  \\
490   &   22:32:54.27& -60:33:20.1  & 23.8 &  0.00 &	0.00  &  0.00  &  0.00 &  22.68 &  0.33 &  0.60  &  0.00   &	0.565		&   24.0  &   0.51  \\
491   &   22:32:54.33& -60:33:20.6  & 23.8 &  0.00 &	0.00  &  0.00  &  0.00 &  22.43 &  0.31 &  0.56  &  0.00   &	0.565		&   23.8  &   0.47  \\
508   &   22:32:45.39& -60:33:23.0  & 23.9 &  23.1 &	0.26  &  0.47  &  0.20 &  0.000 &  0.00 &  0.00  & -1.00   &	0.280$^a$	&   23.9  &   0.26  \\
528   &   22:32:53.63& -60:32:35.9  & 21.1 &  24.1 &	0.09  &  13.0  &  0.59 &  20.66 &  0.54 &  0.64  &  0.02   &	0.365		&   21.0  &   0.90  \\
544   &   22:32:59.43& -60:32:40.1  & 22.7 &  25.2 &	0.07  &  10.5  &  0.10 &  21.67 &  0.60 &  0.81  &  0.01   &	0.480$^a$	&   22.5  &   1.01  \\
549   &   22:32:50.90& -60:32:42.9  & 20.8 &  19.9 &	0.14  &  1.57  &  0.13 &  22.32 &  0.56 &  0.19  &  0.60   &	0.579		&   20.7  &   0.27  \\
575   &   22:32:55.75& -60:32:13.4  & 24.1 &  25.9 &	0.18  &  0.43  &  0.10 &  23.18 &  0.34 &  0.20  &  0.03   &	0.759		&   23.6  &   0.56  \\
577   &   22:32:53.09& -60:32:38.9  & 23.7 &  23.4 &	0.09  &  1.11  &  0.49 &  21.84 &  0.27 &  0.51  &  0.05   &	0.540$^a$	&   23.3  &   0.44  \\
578   &   22:32:52.07& -60:31:41.0  & 23.1 &  23.8 &	0.23  &  1.41  &  0.32 &  22.16 &  0.40 &  0.51  &  0.18   &	0.512		&   22.7  &   0.57  \\
599   &   22:32:50.80& -60:31:41.6  & 23.5 &  0.00 &	0.00  &  0.00  &  0.00 &  22.06 &  0.26 &  0.42  &  0.00   &	0.515		&   23.5  &   0.45  \\
601   &   22:32:54.08& -60:31:42.7  & 22.8 &  23.8 &	0.00  &  10.4  &  0.37 &  20.39 &  0.17 &  0.19  &  0.01   &	0.513		&   22.4  &   0.28  \\
609   &   22:32:57.75& -60:32:33.0  & 21.9 &  26.3 &	0.50  &  10.1  &  0.28 &  19.96 &  0.18 &  0.33  &  0.12   &	0.517		&   21.8  &   0.32  \\
610   &   22:32:57.97& -60:32:34.3  & 21.3 &  21.3 &	0.07  &  2.24  &  0.17 &  21.83 &  0.64 &  0.42  &  0.08   &	0.760		&   21.2  &   0.99  \\
635   &   22:32:56.06& -60:31:48.9  & 21.3 &  25.5 &	1.25  &  11.9  &  0.96 &  21.74 &  0.61 &  0.30  & -0.03   &	0.514		&   21.1  &   1.03  \\
638   &   22:32:52.69& -60:31:53.2  & 24.1 &  0.00 &	0.00  &  0.00  &  0.00 &  21.81 &  0.33 &  0.77  &  0.00   &	0.940$^a$	&   23.8  &   0.50  \\
645   &   22:32:57.13& -60:31:52.2  & 22.3 &  0.00 &	0.00  &  0.00  &  0.00 &  20.05 &  0.23 &  0.36  &  0.00   &	0.600$^a$	&   21.8  &   0.38  \\
663   &   22:32:50.28& -60:32:03.3  & 22.2 &  21.3 &	0.17  &  1.82  &  0.09 &  0.000 &  0.00 &  0.00  &  1.00   &	0.414		&   22.2  &   0.17  \\
673   &   22:32:55.53& -60:32:17.4  & 23.3 &  22.5 &	0.19  &  0.68  &  0.53 &  23.06 &  0.46 &  0.37  &  0.26   &	0.519		&   22.9  &   0.55  \\
676   &   22:32:50.80& -60:31:59.7  & 24.2 &  22.8 &	0.23  &  0.50  &  0.33 &  0.000 &  0.00 &  0.00  & -1.00   &	0.318		&   24.0  &   0.23  \\
684   &   22:32:55.71& -60:32:11.4  & 21.5 &  19.6 &	0.06  &  0.97  &  0.00 &  20.57 &  0.29 &  0.31  &  0.25   &	0.673		&   21.3  &   0.35  \\
687   &   22:32:50.34& -60:32:01.1  & 24.3 &  28.9 &	0.08  &  3.33  &  0.10 &  23.20 &  0.38 &  0.50  &  0.01   &	0.920$^a$	&   24.0  &   0.63  \\
691   &   22:32:48.83& -60:32:03.5  & 24.1 &  22.4 &	0.14  &  1.40  &  0.15 &  0.000 &  0.00 &  0.00  &  1.00   &	0.980$^a$	&   24.0  &   0.14  \\
708   &   22:32:52.70& -60:32:07.2  & 21.8 &  24.7 &	0.76  &  2.63  &  0.45 &  21.91 &  0.47 &  0.28  &  0.29   &	0.464		&   21.5  &   0.78  \\
758   &   22:32:54.77& -60:32:15.4  & 21.6 &  19.1 &	0.06  &  0.75  &  0.11 &  20.39 &  0.20 &  0.19  &  0.36   &	0.480$^a$	&   21.6  &   0.18  \\
759   &   22:32:53.85& -60:32:13.1  & 24.2 &  24.6 &	0.30  &  1.12  &  0.87 &  22.31 &  0.27 &  0.39  &  0.06   &	0.114		&   23.6  &   0.44  \\
763   &   22:33:00.23& -60:32:33.9  & 20.5 &  0.00 &	0.00  &  0.00  &  0.00 &  20.57 &  0.45 &  0.29  &  0.00   &	0.415		&   20.7  &   0.77  \\
766   &   22:32:56.04& -60:32:20.4  & 22.5 &  23.8 &	0.21  &  0.67  &  0.23 &  22.73 &  0.72 &  0.43  &  0.05   &	0.519		&   21.9  &   1.16  \\
768   &   22:32:53.04& -60:32:17.0  & 23.7 &  24.0 &	0.21  &  1.54  &  0.29 &  23.82 &  1.04 &  0.71  &  0.16   &	0.760$^a$	&   22.9  &   1.45  \\
770   &   22:32:48.87& -60:32:16.0  & 23.0 &  25.1 &	0.31  &  2.08  &  0.44 &  21.41 &  0.31 &  0.61  &  0.10   &	0.580		&   22.8  &   0.51  \\
774   &   22:32:52.01& -60:32:14.9  & 23.4 &  0.00 &	0.00  &  0.00  &  0.00 &  22.00 &  0.30 &  0.39  &  0.00   &	0.800$^a$	&   23.2  &   0.52  \\
822   &   22:32:54.78& -60:32:28.9  & 24.1 &  24.4 &	0.11  &  1.51  &  0.10 &  23.70 &  0.41 &  0.06  &  0.08   &	0.760$^a$	&   23.5  &   0.64  \\
10010 &   22:33:05.87& -60:33:43.3  & 22.6 &  22.4 &	0.37  &  0.54  &  0.27 &  0.000 &  0.00 &  0.00  & -1.00   &    0.640$^a$	  &   22.4  &	0.37  \\
\hline

\end{tabular}

\end{minipage}

\end{table*}

For galaxies with z$<$0.4 the best--fitting galaxy parameters in the F606W band
are shown in Table~\ref{data2}. 

\begin{table*}

 \centering
 \begin{minipage}{170mm}
  \caption{F606W Best Morphological Parameter Values  For
  Galaxies with I$_{814}$(AB)$<$24.5 mag and z$<$0.4 in the HDF--N and HDF--S.}
  \begin{tabular}{cccccccccccccc}
  \hline
   (1)  & (2) & (3) & (4)  & (5) & (6)  & (7) & (8)  & (9)
   & (10) & (11) & (12) & (13)       \\ 
  ID  &V$_{606}$(cat) & $\mu_{e,b}$ & r$_{e,b}$ & n & $\epsilon_b$ &
  $\mu_d$(0) & h & $\epsilon_d$ & B/T & z  & V$_{814}$(mod) &
  r$_e$  \\
 &  mag  & mag/''$^2$ & ('')  & &  &
  mag/''$^2$ & ('') &  &  &    & mag &
    ('')    \\ 
     \hline
&      &  &   & & & HDF--N &
  &  &  &  &      &
   &     \\  
 \hline
10     & 24.4 &  26.1 &   0.11  &  1.35  &  0.26 &  23.6 &  0.44 &  0.40  &	0.02	&	0.080$^a$   &	23.9  &   0.72  \\
11     & 23.6 &  0.00 &   0.00  &  0.00  &  0.00 &  22.1 &  0.33 &  0.60  &	0.00	&	0.321	    &	23.6  &   0.50  \\
55     & 23.9 &  0.00 &   0.00  &  0.00  &  0.00 &  23.3 &  0.53 &  0.75  &	0.00	&	0.200$^a$   &	24.2  &   0.75  \\
85     & 20.1 &  23.1 &   0.28  &  1.52  &  0.10 &  20.7 &  0.70 &  0.40  &	0.06	&	0.089	    &	20.0  &   1.10  \\
191    & 24.8 &  25.5 &   0.37  &  4.40  &  0.24 &  0.00 &  0.00 &  0.00  &	1.00	&	0.372	    &	24.6  &   0.37  \\
350    & 21.8 &  24.7 &   0.15  &  2.58  &  0.55 &  21.9 &  0.53 &  0.25  &	0.02	&	0.320	    &	21.6  &   0.88  \\
351    & 24.5 &  27.3 &   0.32  &  2.10  &  0.30 &  23.7 &  0.46 &  0.54  &	0.06	&	0.240$^a$   &	24.2  &   0.74  \\
363    & 24.9 &  0.00 &   0.00  &  0.00  &  0.00 &  22.1 &  0.18 &  0.60  &	0.00	&	0.010$^a$   &	24.8  &   0.28  \\
445    & 22.6 &  21.7 &   0.22  &  0.87  &  0.53 &  23.4 &  0.60 &  0.56  &	0.56	&	   0.319       &   22.5  &   0.37  \\
537    & 21.5 &  21.5 &   0.24  &  1.03  &  0.42 &  22.4 &  0.52 &  0.04  &	0.36	&	   0.321       &   21.4  &   0.55  \\
611    & 22.8 &  0.00 &   0.00  &  0.00  &  0.00 &  20.5 &  0.28 &  0.70  &	0.00	&	   0.319       &   22.6  &   0.48  \\
619    & 24.0 &  25.9 &   0.72  &  4.00  &  0.23 &  0.00 &  0.00 &  0.00  &	1.00	&	 0.370       &   23.9  &   0.72  \\
631    & 24.0 &  22.6 &   0.22  &  0.37  &  0.62 &  23.4 &  0.25 &  0.25  &	-0.52	&	 0.321       &   23.9  &   0.22  \\
637    & 23.4 &  26.3 &   1.66  &  1.00  &  0.05 &  0.00 &  0.00 &  0.00  &	-1.00	&	 0.089       &   22.6  &   1.66  \\
669    & 24.1 &  0.00 &   0.00  &  0.00  &  0.00 &  22.6 &  0.27 &  0.29  &	0.00	&	0.010$^a$   &	23.9  &   0.39  \\
757    & 19.9 &  21.7 &   0.18  &  0.42  &  0.62 &  21.3 &  1.09 &  0.51  &	0.02	&	   0.199       &   19.9  &   1.79  \\
869    & 21.5 &  24.1 &   0.23  &  1.62  &  0.67 &  22.4 &  0.64 &  0.37  &	0.03	&	   0.321       &   21.8  &   1.04  \\
914    & 18.7 &  20.3 &   0.63  &  2.30  &  0.23 &  0.00 &  0.00 &  0.00  &	1.00	&	   0.089       &   18.5  &   0.63  \\
1029   & 24.3 &  23.2 &   0.21  &  1.86  &  0.40 &  0.00 &  0.00 &  0.00  &	1.00	&	   0.130       &   24.1  &   0.21  \\
1067   & 19.5 &  20.0 &   0.11  &  0.51  &  0.18 &  19.2 &  0.42 &  0.34  &	0.06	&	   0.139       &   19.5  &   0.66  \\
     \hline
&  &  &  &     & & HDF--S &  &  &  &     &  &   &     \\  
\hline
184  & 20.8 &  23.6 &	0.19  &  6.03  &   0.32 &  20.7 &  0.41 &  0.27 &    -0.08    & 0.340	  &   20.9  &  0.66 \\
186  & 20.8 &  23.7 &	0.77  &  6.84  &   0.04 &  0.00 &  0.00 &  0.00 &     1.00    & 0.328	  &   20.6  &  0.77 \\
244  & 21.7 &  0.00 &	0.00  &  0.00  &   0.00 &  21.0 &  0.67 &  0.80 &     0.00    & 0.173	  &   21.7  &  1.00 \\
283  & 24.4 &  27.2 &	0.24  &  3.11  &   0.12 &  23.6 &  0.59 &  0.55 &     0.04    & 0.180$^a$ &   23.6  &  0.96  \\
314  & 23.5 &  0.00 &	0.00  &  0.00  &   0.00 &  21.6 &  0.18 &  0.14 &     0.00    & 0.220$^a$ &   23.5  &  0.28  \\
438  & 25.1 &  0.00 &	0.00  &  0.00  &   0.00 &  22.6 &  0.23 &  0.55 &     0.00    & 0.359	  &   24.7  &  0.37 \\
446  & 21.9 &  24.0 &	0.12  &  12.1  &   0.09 &  21.2 &  0.23 &  0.44 &     0.19    & 0.364	  &   22.8  &  0.37 \\
487  & 24.2 &  24.2 &	0.43  &  0.69  &   0.14 &  0.00 &  0.00 &  0.00 &    -1.00    & 0.240$^a$ &   23.6  &  0.43  \\
508  & 24.0 &  24.3 &	0.37  &  0.64  &   0.51 &  0.00 &  0.00 &  0.00 &    -1.00    & 0.280$^a$ &   24.8  &  0.37  \\
528  & 21.7 &  0.00 &	0.00  &  0.00  &   0.00 &  21.1 &  0.45 &  0.56 &     0.00    & 0.365	  &   21.8  &  0.68  \\
676  & 24.5 &  23.4 &	0.25  &  1.42  &   0.39 &  0.00 &  0.00 &  0.00 &    -1.00    & 0.318	  &   24.1  &  0.25  \\
759  & 24.4 &  23.8 &	0.18  &  0.36  &   0.26 &  23.0 &  0.31 &  0.48 &     0.17    & 0.114	  &   24.0  &  0.42  \\
\hline
\label{data2}
\end{tabular}

\end{minipage}

\end{table*}

Our code has already been used and tested in a variety of HST and ground--based
studies of  nearby and distant galaxies: the quantitative morphological
analysis of  galaxies in Abell 2443 (Trujillo et al. 2001a) and Coma
(Guti\'errez et al. 2004; Aguerri et al. 2004) clusters, the number density
evolution of galaxies in the SSA13 and SSA22 Hawaiian Deep Fields (Aguerri \&
Trujillo 2002), the ISAAC near--infrared  images of the HDF--S (Trujillo et al.
2004) and the Flanking Fields HST images (Graham et al. 2004). In Trujillo et
al. (2004) we have also made a comparison between our code and the GALFIT code
(Peng et al. 2002) finding an excellent agreement.

\begin{figure}

\vskip 6.0cm
{\includegraphics{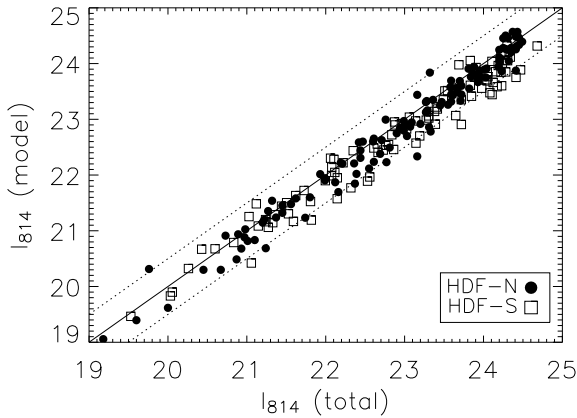}}

  \caption{The total magnitude retrieved from the fitting model is compared to
  the total magnitude measured in a model independent way (catalogs' magnitude)
  for the galaxies analysed in this paper. Dashed lines corresponds to $\pm$0.5
  mag over the assumption of an exact agreement.}

\label{magcompara}
  \end{figure}

A preliminary way of examining the goodness of our structural parameter
determination is to compare the best fit model magnitude with the magnitude
measured in a model independent way (Fig.~ \ref{magcompara}). The model
independent magnitude is the catalogs' magnitude. As expected from its
extrapolation to the infinity, the model magnitude is almost always equal or
brighter than the model independent determination.

\section{Object Morphological Classification}

Following previous works (e.g. MS98) our  galaxy classification scheme relies
on  B/T (Simien \& de Vaucouleurs 1986). Galaxies are considered
``ellipticals'' if their B/T$>$0.6. If this is the situation the galaxies are
reanalysed with just a S\'ersic component and the parameters provided here
correspond to this case. Galaxies with B/T between 0.5--0.6 are classified as
S0 and objects with B/T$<$0.5 are named ``spirals''. However, there is a group
of objects that can not be well fitted by our models. These objects have
peculiar shapes (and consequently, their isophotes are hardly fitted by
ellipses) and are classified  in the Irr/Merger category. These objects are
indicated in Tables \ref{data} and \ref{data2} with B/T=-1. In addition, there
are a group of  visually classified spiral galaxies where a reliable fit is not
achieved due to the presence of strong secondary structures as bars, spiral
arms, etc. These galaxies are indicated with a value of B/T$<$0.


The results of our galaxy classification are shown in Table \ref{percentage}.
There are some discrepancies between the number of galaxies classified as E/S0
and Sa/Sc in the HDF--S and HDF--N. These differences illustrate  the
strong field--to--field variations present in small observing  volumes.

\begin{table}
 \centering
 \begin{minipage}{140mm}
  \caption{Coarse binning of galaxy classifications.}
  \begin{tabular}{rrrrr}
  \hline
   Type     &  B/T &  HDF--N        &  HDF--S  & Total\\
  \hline
 E/S0       & $>$0.5 & 20\% &  12\%  & 16\% \\
 Sa/Sc      & $<$0.5 & 63\% &  74\%  & 68\% \\
Irr/Merger  &        & 17\% &  14\%  & 16\% \\
\hline
\label{percentage}
\end{tabular}
\end{minipage}
\end{table}

\section[]{Determination of Accuracy}

To calibrate the accuracy of our parameter determination we have made two
different tests. First, we have run our code on simulated  artificial galaxies.
This permits an evaluation of the apparent magnitude limit cut--off. Galaxies
brighter than this magnitude limit are then claimed to be recovered with a
certain degree of accuracy. We also have made a direct comparison
between the structural parameters recovered using our algorithm and those
recovered using GIM2D on the HDF--N (MS98). We describe these two issues in the
following subsections.

\subsection{Simulations}

Simulations are a key element in the interpretation of the observations. For
fainter objects (i.e. at decreasing the signal--to--noise S/N) the structural
parameters of the galaxies are progressively more difficult to determine. To
quantify this effect it is necessary to run realistic simulations matching the
observational conditions. We simulated two kind of objects. First, we created
bulge--only (i.e. pure elliptical) artificial galaxies, and second, we
constructed bulge+disc galaxies. In both cases, we made 150 artificial objects
with structural parameters randomly distributed in the following ranges:

\begin{itemize}

\item  bulge--only structures: 19$\leq$I(AB)$\leq$25, 0.05$''$$\leq$$r_{e}
\leq$0.6$''$, 0.5$\leq n \leq 4$, and $0\leq \epsilon \leq 0.6$ 

\item  \ bulge+disc \ structures: 19$\leq$I(AB)$\leq$25, 0.05$''$$\leq$$r_{e}
\leq$0.6$''$, 0.5$\leq n \leq 4$, and $0\leq \epsilon_{b} \leq 0.4$, 0.2''$\leq
h \leq 1.5 ''$, 0$\leq B/T \leq 1$, and $0\leq \epsilon_{d} \leq 0.6$. 

\end{itemize}

Artificial galaxies were created by using the IRAF task MKOBJECT. We support
as an input to this task the surface brightness distribution coming from our
detailed convolution between the PSF and the original model.  To simulate the
real conditions of our observations, we added a background sky image (free of
sources) taken from a piece of the real image. The PSF FWHM in the simulation
was set at 0.147$''$  and assumed  known exactly. The pixel scale of the
simulation was 0.04$''$, as are the drizzled HDF pixel scales. The same procedure
was used to process both the simulated and  actual data.

From these simulations we find that the bulge and disc parameters (as well as
the galaxy global properties as total magnitude, r$_e$, etc.) can be determined
with an accuracy of $\sim$10\% for galaxies with  I$_{814}$(AB)$<$24.5 mag. 
Consequently, we consider I$_{814}$(AB)$=$24.5 mag as our limiting magnitude
for a reliable determination of the parameters. Further details of how the
simulations are constructed and the errors determined can be found in Trujillo
et al. (2001a) and Aguerri \& Trujillo (2002).

\subsection{Comparison with Previous Works}

MS98 using a different code, GIM2D (Simard 1998), conducted a quantitative
morphological analysis  on the HDF--N close to that presented here. The
main differences between both analyses are: the limiting apparent magnitude and
the model for the bulge component. GIM2D was  run down to I$_{814}$(AB)$<$26
mag, i.e. 1.5 mag deeper than in the present analysis. However, although
individual errors for the galaxy parameters are available, there are not
simulations showing  the accuracy of the structural parameters  as a function
of the apparent magnitude. Another difference is that GIM2D was run using a de
Vaucouleurs model ($n$=4 fixed) for the bulges. Fixing $n$ to 4 has the
advantage of decreasing the number of free parameters, however, in the nearby
universe  a range of $n$ (from 0.3 to 6) is found (Andredakis, Peletier \&
Balcells 1995, Graham 2001). Moreover, $n$ relates with the total magnitude of
the bulge component and it would be of interest to know if this relation
evolves with time (Aguerri \& Trujillo 2004).

\begin{figure}

\vskip 6.0cm
{\includegraphics{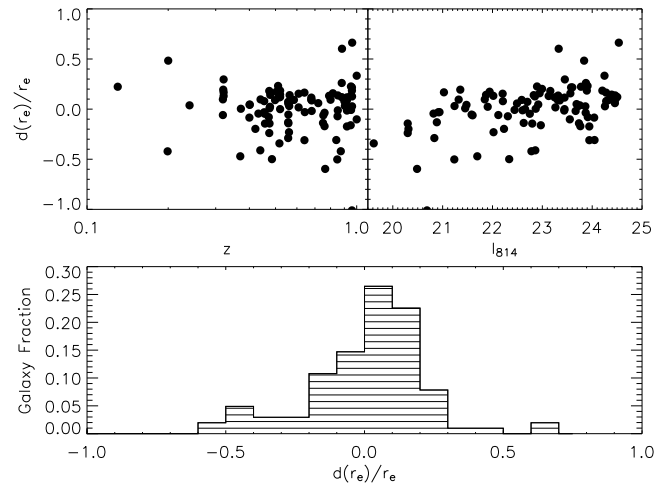}}

  \caption{The relative error between the size estimation using our code and
GIM2D
d(r$_e$)/r$_e$=2$\times$(r$_e$(GIM2D)-r$_e$(ours))/(r$_e$(GIM2D)+r$_e$(ours))
is shown versus $z$ and versus the apparent I$_{814}$ magnitude. The histogram
shows that for $\sim$75\% of the galaxies the difference on size is less than
20\%.}

\label{recompara}
  \end{figure}

In Fig. \ref{recompara} we show the comparison between the global size of the
galaxies r$_e$ as obtained from our code ($n$ free) and GIM2D ($n$=4) versus
$z$ and versus the apparent magnitude. The comparison is limited to
I$_{814}$(AB)$<$24.5 mag and z$<$1. There is a good agreement between these two
different algorithms ($\sim$75\% of the galaxies have a size difference less
than 20\%). However, we find a small offset of 5\% toward smaller r$_e$ in our
code relative to GIM2D. This offset seems to be real in the sense that a
T-Student test rejects the hypothesis that this is produced by chance by more
than 99.9\%. This offset is not dependent on $z$ or magnitude and (due to it is
not present in our simulations) must be intrinsically related to the way our
code and GIM2D estimate the global galaxy size. As explained before our code
uses Eq. (5) to estimate r$_e$. This equation is derived by the definition of
effective radius (i.e. L(r$_e$)=0.5L($\infty$)). On the other hand, GIM2D
evaluates r$_e$ by integrating the bulge and disc expressions out to infinity
and then evaluating r$_e$. It is possible that their assumption of $n$=4 plays
here this minor role including more light than what it is actually observed
with a $n$ smaller. If this is the case we would expect that their size would
be slightly larger than ours. In any case, it is important to note that this
offset is much smaller than the standard deviation found at comparing both
codes ($\sim$20\%).

\begin{figure}

\vskip 6.0cm
{\includegraphics{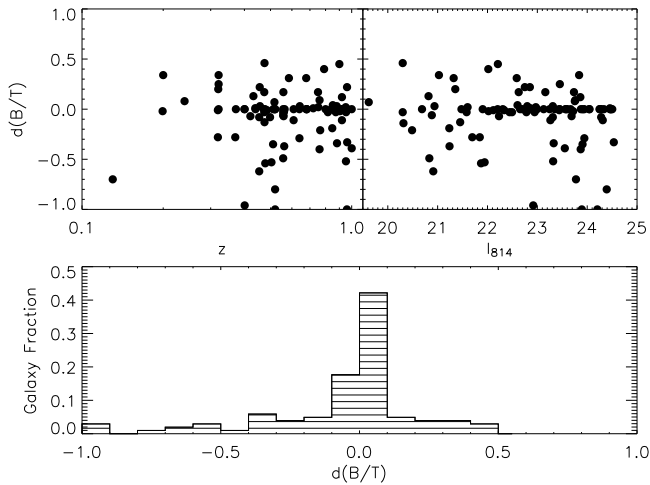}}

  \caption{The difference between the B/T estimation using our code and GIM2D
d(B/T)=B/T(GIM2D)-B/T(ours) is shown versus $z$ and versus the apparent
I$_{814}$ magnitude. The histogram shows that for $\sim$73\% of the galaxies
 B/T differs less than 0.2 units.}

\label{btcompara}
  \end{figure}

 We also present the comparison between our B/T ratio and GIM2D in Fig.
\ref{btcompara}. The global properties of the galaxies r$_e$ and B/T are shown
to be very robust for this sample of galaxies. In the nearby Universe, Simien
\& de Vaucouleurs (1986), using a $n$=4 bulge, did an estimation of the B/T
ratio for a sample of galaxies ranging a large variation of morphological types. 
Graham (2001) conducted a similar analysis using this time  a $n$ free bulge. 
He found that the use of a $n$=4 bulge (instead of a $n$ free) produces an
overestimation of the B/T ratio. We do not see a strong difference between our
B/T values and the GIM2D estimations. This is probably due to most of our
galaxies having small B/T ratios and consequently the bulge component is not
dominant. However, the use of a different model for the bulge will have
consequences when dealing with bulge properties such as  r$_{e,b}$.

\begin{figure}

\vskip 6.0cm
{\includegraphics{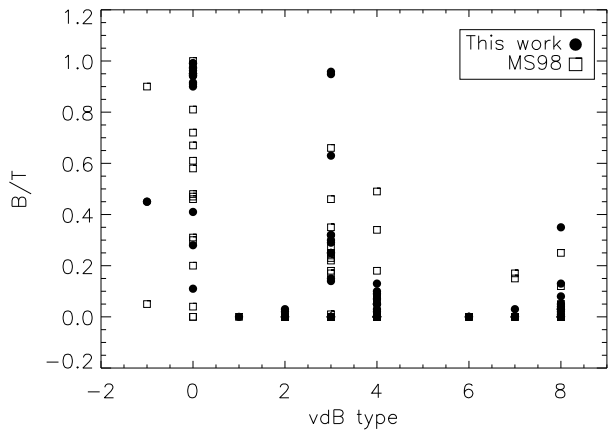}}

  \caption{Comparison of the parameter B/T with the vdB morphological type 
  derived from the visual classification of van den Bergh et al. (1996).}

\label{bttype}
  \end{figure}

We also compare our B/T determination with the morphological type obtained from
van den Bergh et al. (1996, hereafter vdB96). This is shown in Fig.
\ref{bttype}.  The numerical classification used by vdB96 (Abraham et al. 1996)
is as follows: E/star: -1, E: 0, E/S0/Sa: 1, S0/Sa: 2, Sa/Sab: 3, Sb/S/Ir: 4,
Sc: 5, Ir: 6, Peculiar/Mergers: 7 or 8. As  explained before, galaxies with
B/T$>$0.6 were reanalysed and fitted with a bulge--only S\'ersic component
(i.e. B/T=1). For illustrative purpose, to avoid the appearance of just one
point at B/T=1 in Fig.~ \ref{bttype}, we have randomly moved all the galaxies
with this B/T value a 5\% around 0.95. The main difference between our analysis
and MS98 is  the group of galaxies classified by vdB96 as vdB=0. Contrary to
MS98, most of the vdB96 visual ellipticals are also classified as B/T=1 in our
quantitative analysis.

 The fractions of the different morphological types according to vdB96 (down to
I$_{814}$(AB)$<$25 mag and not $z$ restrictions) are: 30\% E/S0, 31\% S/Irr and
39\% not classified. However, these percentages  depend on the limiting
magnitude   and the limiting redshift used. To make a reasonable comparison
between the different morphological types which follow from the three different
methods (ours, MS98 and vdB96) we have used exactly the same galaxies. Our
results are presented on Table \ref{percompara}. As we can see the agreement
between the different methods is  good. Consequently, previous claimed
discrepancies between visually and quantitative methods (MS98) are shown here
to be potentially due to different limiting magnitude criteria and redshift.
However, although the fractions may be similar for each study, different
galaxies may populate each bin. We have checked if this is the case by
comparing galaxy by  galaxy and found that the agreement in each bin is greater
than 65\%.

\begin{table}
 \centering
 \begin{minipage}{70mm}
  \caption{Coarse comparison of galaxy classifications in the HDF--N down
   to I$_{814}$(AB)$<$24.5 mag and z$<$1 for the same group of galaxies using
   three different methods. The number of analysed galaxies 
   is 102.}
  \begin{tabular}{rrrrr}
  \hline
   Type     &  B/T & This work  &  MS98  & vdB96 \\
  \hline
 E/S0       & $>$0.5 & 20\% &  15\%  & 22\% \\
 S/Irr      & $<$0.5 & 80\% &  85\%  & 78\% \\
\hline
\label{percompara}
\end{tabular}
\end{minipage}
\end{table}

We have also undertaken a comparison with the analysis done in the HDF--S by
Moth \& Elston (2002).  In that work, 33 galaxies with 0.5$\leq$z$\leq$1.2 and
I$_{814}$$<$25 are analysed. The model used for extracting structural
information is a single S\'ersic law and not a bulge+disc decomposition. For
that reason the comparison is not straightfoward. For the 20 galaxies we have
in common, we find a very good agreement in the global size (difference less
than 20\%) for 18 objects (90\%).  A similar analysis using also a single
S\'ersic law is  done in Trujillo et al. (2004) for K$_s$--band selected
galaxies down to K$_s$=23.5. Trujillo et al. analysed in the I$_{814}$--band
galaxies up to $z$$\sim$0.8. The agreement in the global sizes for the common
61 galaxies is good, with $\sim$68\% of the objects having a difference less
than 10\% .

\section[]{Completeness and Redshift-dependent selection effects}

The ability to detect a galaxy is critically dependent on its apparent surface
brightness. Galaxies with larger B/T are easier to detect because they are more
concentrated, and galaxies with larger radii are harder to detect than smaller
ones at given total magnitude. Therefore, it is of interest to know whether our
observed parameter distribution would be biased by the galaxy selection
function of the SExtractor detection algorithm (used in both catalogs we are
using). The evaluation of completeness of the HDF--N  was carefully model in
detail by MS98 (their Fig. 5). According to these authors galaxies brighter
than I$_{814}$(AB)$<$24.5 are fully recovered. Due to the very similar
characteristics of the HDF--S these results are also  valid for that field.
This means that in the sample of galaxies shown in this work the incompleteness
is not playing any role and consequently we will not take it into account in
the subsequent analysis.

\begin{figure}

\vskip 6.0cm
{\includegraphics{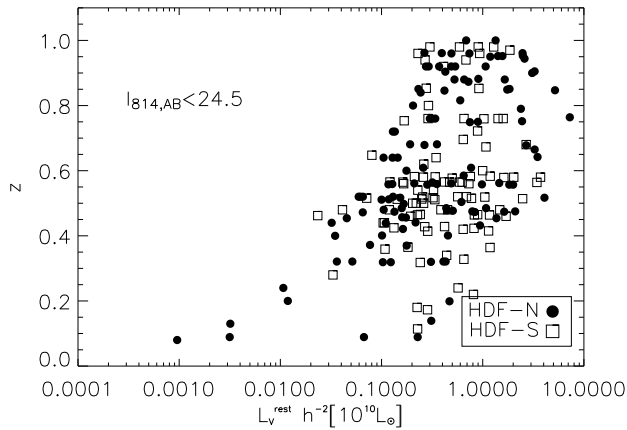}}

  \caption{The L$_V$-z diagram for the objects selected in our sample with
  I$_{814}$(AB)$<$24.5 mag. }

\label{zlum}
  \end{figure}

Redshift--dependent observational biases can also mimic real evolutionary
changes in the galaxy population (Simard et al. 1999). For a given magnitude
limit (in our case I$_{814}$(AB)=24.5 mag) there is a corresponding threshold
in the observable restframe luminosity which increases with redshift. We
illustrate this in Fig.~\ref{zlum}. This figure shows the rest--frame
luminosity in the V--band versus the redshift.

To estimate the rest--frame luminosity in the V--band we make use of the
complete filter coverage of the HDFs. We have used the fluxes measured in the
U$_{300}$, B$_{450}$, V$_{606}$, I$_{814}$, J, H and K bands provided by the
Fern\'andez--Soto et al. (1999; HDF--N) and Labb\'e et al. (2003; HDF--S)
catalogs.  With these fluxes we construct observed SEDs for every galaxy. As we
have redshift data for our sources we estimate the redshifted wavelength of the
center of the local V-band filter ($\lambda$=5510 \AA) (i.e. we evaluate 
$\lambda$=5510$\times$(1+z)). At that wavelength, we measure the  restframe
V-band flux by linearly interpolating the catalog's fluxes at the two closest
$\lambda$ values. The advantange of using this method is that we do not rely on
any SED model to estimate the luminosity. There are two main sources of
uncertainty:
\begin{enumerate}
 \item The uncertainty due to the  photometric redshifts: most
of our galaxies  (70\% in the HDF--N and 60\% in the HDF--S) have z
spectroscopically determined, however, the remaining fractions have an
uncertainty of dz/(1+z)$\sim$0.1. This uncertainty in $z$ can be translated into
luminosity uncertainty  of $\sim$30\%.
 \item The uncertainty due to
the total flux calibration: the reliability of the total flux measured through a
given filter will depend on the S/N of the sources. Fortunately, the HDFs are
so deep, and the objects selected in this work so bright, that this effect does not
play a major role. In fact, we can see from the comparison between our model
magnitudes (which are an extrapolation of the total magnitude up to an infinite
radial extension) and the aperture magnitude (our Fig. \ref{magcompara}) that
the mean discrepancies between both is $\sim$0.2 mag, being the ``true'' flux
value probably in between these two estimates. The above uncertainty in the
magnitude is translated into a luminosity uncertainty  of $\sim$20\%.
\end{enumerate}

According to the above two points we will assume on what follows an uncertainty
in the restframe V--band luminosities of $\sim$35\%.

In Fig.~ \ref{zlum} we show how our high--z sample represents only the most
luminous fraction of the galaxy population. The lack of bright galaxies at low
redshift results directly from their low volume densities coupled with the
small volume of the HDF over this redshift range. Galaxies with L$_V>2\times10^9$
h$^{-2}$ L$_\odot$ (i.e. M$_B\le$-17.5 assuming B-V=0.9) can be
studied along the full redshift range.

\section[]{HDF--N and HDF--S luminosity--size relations out to $z$$\sim$1}

\begin{figure}

\vskip 6.0cm
{\includegraphics{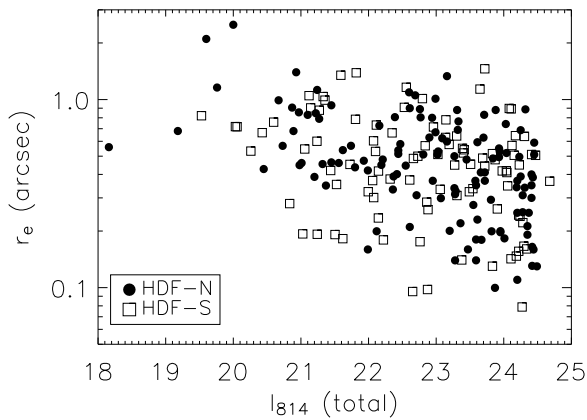}}

  \caption{The relation between the model--independent global magnitude versus
the apparent global semi--major effective radius size (in the
I$_{814}$--band).}

\label{magre}
  \end{figure}

It is of interest to know whether the population of galaxies in the HDF--N and
HDF--S present similar global structural properties. We show in Fig.~
\ref{magre} the relation between the model--independent global magnitude versus
the apparent global semi--major effective radius size (in the I$_{814}$--band).
We have run the generalization of the Kolmogorov--Smirnov (K--S) test on two
dimensional distributions (Fasano \& Franceschini 1987) and found that the null
hypothesis (i.e. that the two distribution have a similar origin) can not be
rejected.

\begin{figure}

\vskip 6.0cm
{\includegraphics{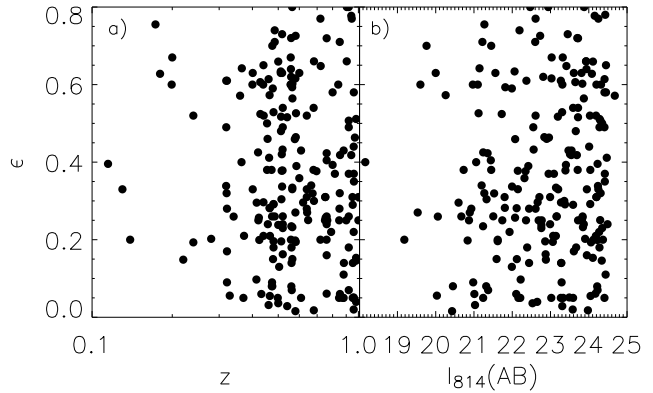}}

  \caption{The intrinsic  (i.e. the recovered non--PSF affected) ellipticity
of the galaxies versus: a) the redshift of the observed sources and b) the
apparent I$_{814}$ total magnitudes. No clear relation is observed.}

\label{ellipticity}
\end{figure}

 In what follows, to facilitate the comparison with local
calibrating data, we use, instead of the  global semi--major effective radius
r$_e$, the {\it circularized effective radius} r$_{e,c}$ (i.e. we use
$r_{e,c}=r_e\sqrt{(1-\epsilon)}$ with $\epsilon$ the intrinsic, non--PSF
affected, projected ellipticity of the galaxy). For those galaxies where
$B/T$$<$1, $\epsilon=\epsilon_d$.  To be sure that the estimate of the
intrinsic ellipticity of our sources, and hence the conversion to circularized
effective radius, was not systematically affected by the PSF effect we
studied the presence of any obvious trend with z or the I$_{814}$ apparent
magnitude (Fig. \ref{ellipticity}).  No clear relation is found.

\begin{figure}

\vskip 6.0cm
{\includegraphics{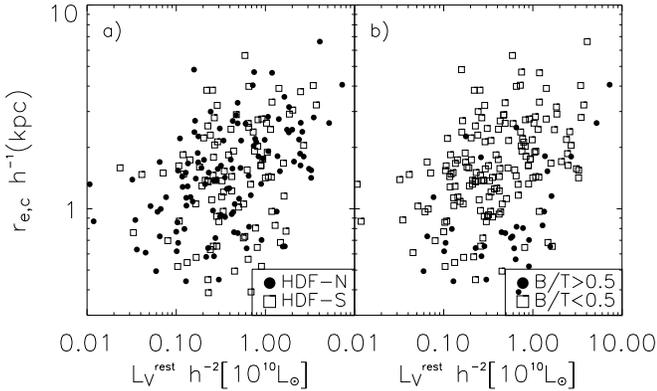}}

  \caption{a) The distribution of circularized rest--frame  sizes versus the
  rest--frame V--band luminosities is shown separating  galaxies according to
  their ownership to the HDF--N or to the HDF--S. b) Same as a) but this time
  segregating the points between late--type (B/T$<$0.5) and early--type
  (B/T$>$0.5) galaxies. For clarity error bars are not shown. The mean size
  relative error is 20\% and the mean luminosity error 30\%.}

\label{relum}
  \end{figure}

 We have conducted a similar analysis to that presented in Fig.~\ref{magre}
using the rest--frame properties of the above galaxies. This is shown in Fig.~
\ref{relum}a. In this case, the null hypothesis also can not be rejected.  To
avoid any bias due to K--corrections we have used for galaxies with $z<$0.4 the
structural parameters obtained in the F606W filter (Table~\ref{data2}). For
galaxies with 0.4$<z<$1 the filter used was  F814W. Therefore the sizes are
computed for roughly the same rest--frame wavelenghts (between B and V) over
the entire redshift range. This avoids any  band--dependent size effects. This
would be particularly important in late--type galaxies. In the local universe,
Graham (2001) finds (assuming that the effective radius of the bulge is not
modified) that the scale of the discs increases 15--25\% from R to B band.

 In Fig. ~ \ref{relum}b we show the same sample but making a separation between
early--type  (B/T$>$0.5) and  late--type (B/T$<$0.5) galaxies. At a fixed
luminosity, late--type galaxies are larger than the early--types.

\subsection{Testing the hypothesis of evolution}

\begin{figure}

\vskip 6.0cm
{\includegraphics{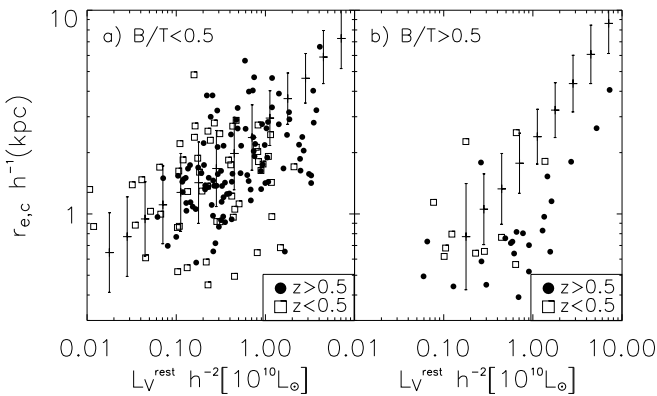}}

  \caption{a) The distribution of rest--frame  sizes versus the rest--frame
  V--band luminosities is shown for late--type (B/T$<$0.5) galaxies. b) Same as
  a) with early--type (B/T$>$0.5) galaxies. The points are separated according
  to the $z$ value. For clarity error bars are not shown. The mean size
  relative error is 20\% and the mean luminosity error 30\%.}

\label{relumbt}
  \end{figure}

Shen et al. (2003) have shown the median and dispersion of the distribution of
the circularized half--light radius (Blanton et al. 2003) for different bands
as a function of the luminosity in the Sloan Digital Sky Survey (SDSS, York et
al. 2000). Their analysis is based on an unprecedented large database of nearby
galaxies with typical z$\sim$0.1. In the following, we closely match the
analysis presented in Trujillo et al. (2004). We use the SDSS g--band (the
closest available filter to our V--band) size distributions (S.  Shen, 2003,
private communication) as a local reference of the size distribution of
galaxies in the nearby universe. The median and dispersion of the Shen et al.
distribution in the g--band separating into early and late type  are shown in
Fig.~ \ref{relumbt}. We also overplot our HDFs data.

The criteria followed to separate between early and late types in the presented
SDSS relations is different from the one we have used. In the Shen et al.
distributions a galaxy is separated into early or late type according to whether
its global S\'ersic index $n$ is larger or smaller than 2.5. The index $n$
derived for these authors corresponds to a global S\'ersic fit to the whole
galaxy, whereas we have used a bulge+disc decomposition and our index $n$
refers only to the bulge component. The separation between early and late type
according to the global index $n$ can be affected by the inclusion of some Sa
galaxies as early--type galaxies. This effect is well illustrated by
Ravindranath et al. (2004, their Fig. 1). Also, some contamination in the
late--type galaxies at lower luminosities is expected from dwarf ellipticals
which have $n$$\sim$1.

\subsubsection{Late--Type galaxies}

 At a given luminosity,  Shen et al. propose for nearby
galaxies the following size log--normal distribution with median $\bar{r}_e$
and logarithmic dispersion $\sigma$:

\begin{equation}
f(r_e|\bar{r}_e(L),\sigma(L))=\frac{1}{\sqrt{2\pi}\sigma(L)}
 e^{-\frac{\ln^2 (r_e/\bar{r}_e(L))}{2\sigma^2(L)}}\frac{dr_e}{r_e}.
\label{sdss}
\end{equation}

To address the problem of size evolution, we have assumed that the functional
form of the size log--normal distribution function (Eq. \ref{sdss}) found for
nearby galaxies also holds at all redshifts, but with evolving parameters:
\begin{eqnarray} \bar{r}_e(L,z)=\bar{r}_e(L,0)(1+z)^{-\alpha} \label{sdss3} \\
\sigma(L,z)=\sigma(L,0)(1+z)^{\beta}. \label{sdss4} \end{eqnarray} Here,
$\bar{r}_e(L,0)$ and $\sigma(L,0)$ are the median size and dispersion provided
at z=0 by the Shen et al. (2003) data, and $\alpha$ and $\beta$ describe the
redshift evolution. Implicit in Eq. \ref{sdss3} and Eq \ref{sdss4} we assume
the same size evolution for all the galaxies independently of their luminosity.
However, in order to  assure that we observe the same type
of galaxies  along the explored redshift range 0$<$$z$$<$1 (see Fig.
\ref{zlum}) we select only galaxies with L$_V$$>$2$\times$10$^9$ h$^{-2}$
L$_\odot$. To explore in much detail the evolution of the luminosity--size
relation with $z$ we have separated our sample in galaxies with $z$$<$0.5 (mean
$z$$\sim$0.4) and galaxies with $z$$>$0.5 (mean $z$$\sim$0.7). The number of
galaxies with L$_V$$>$2$\times$10$^9$ h$^{-2}$ L$_\odot$  is 31 in the lower z
bin and 85 in the higher. Due to the scarce number of objects in the lower z
bin we will concentrate our analysis on the galaxies with z$>$0.5. In addition,
there is a group of outliers with z$<$0.5, luminosities between
0.1--1$\times$10$^{10}$ h$^{-2}$ L$_\odot$ and small sizes that  affects our
analysis at lower redshifts. These galaxies have all z photometrically
determined. The option of taking only galaxies with z spectroscopically
determined for this bin is not useful because of the few number of galaxies
then available.

We evaluate $\alpha$ and $\beta$  using a Maximum Likelihood Method. The
likelihood to maximize is given by:

\begin{equation} L=\prod_{i=1}^n f\otimes g_L\otimes g_{r_e} \end{equation}
Each member (one per galaxy measured) of the product  is a convolution between
the expression $f$ given in Eq. \ref{sdss} and a gaussian representing the
error on the luminosity $g_L$ and another gaussian representing the error on
the size $g_{r_e}$. The width of each gaussian is given by the error at
estimating the luminosity and the size respectively.

\begin{figure}

\vskip 6.0cm
{\includegraphics{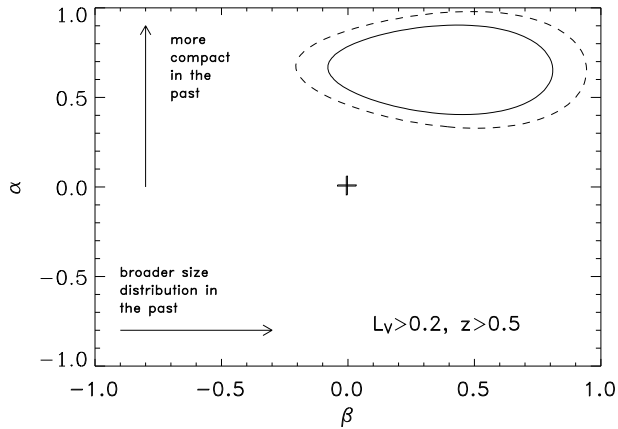}}

  \caption{Luminosity--size evolution for late--type galaxies with z$>$0.5. Likelihood
contours representing the evolution in size and dispersion in the
$\alpha$--$\beta$ plane. Solid line represents the 1$\sigma$ confidence level
contour, dashed line the 2$\sigma$ confidence level. The cross shows the
position of no--evolution in this plane.  Positive values of $\alpha$ represent
decreasing values of the size with redshift. Positive values of $\beta$
represent increasing the intrinsic dispersion $\sigma$ of the population with
redshift. The luminosities are given in 10$^{10}$ L$_\odot$.}

\label{likelihood}
  \end{figure}

The result of this analysis is shown in Fig. ~\ref{likelihood}. According to
this analysis the luminosity--size distribution of disc galaxies presents a
slightly (although compatible with cero) broader dispersion at higher redshift.
The dispersion increases by  $\sim20(\pm25)\%$.  To know whether this increase
in the dispersion is real or not will require a large number of galaxies.

We  observe  a moderate decrease in the sizes $\sim30(\pm10)\%$ at z$>$0.5. In
the luminosity--size distribution a decrease in the size can also be 
interpreted as an increase in the luminosity of the objects. In fact,  this
seems to be a plausible hypothesis to explain the observed
luminosity--size evolution found at high--z. According to the studies of the
evolution of the stellar mass--to--light ratio  (see e.g. Rudnick et al. 2003)
the stellar populations were younger at high--z, and for that reason, the
mass--to--light ratio is a decreasing function with z. Consequently, even
though the size of the galaxies does not decrease with z, the luminosity
evolution of the objects will simulate this effect. Consequently, our observed
evolution in the luminosity--size distribution can be alternatively interpreted
as a decrease in the central surface brightness of $\sim0.77(\pm0.30)$ mag at
at z$\sim$0.7.

The results presented here are in  agreement with those found by Lilly et
al. (1998), Simard et al. (1999) and Ravindranath et al. (2004). These authors
do not find a great change from local galaxies out to $z$=0.7 and a hint of
change at $z$$>$0.7 with shorter objects than those locally observed for a
similar magnitude range. 

\subsubsection{Early--Type galaxies}

\begin{figure}

\vskip 6.0cm
{\includegraphics{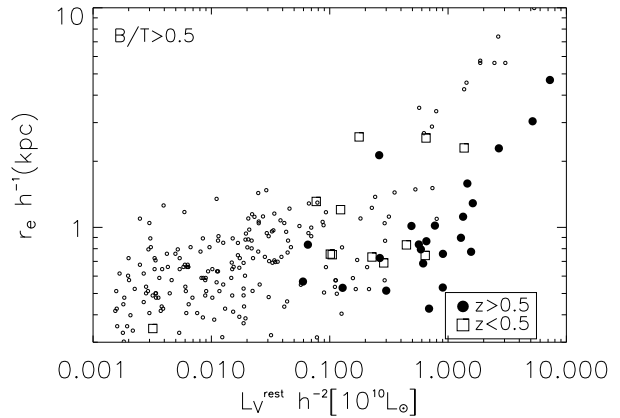}}

\caption{The distribution of rest--frame  sizes versus the rest--frame V--band
luminosities is shown for  early--type (B/T$>$0.5) galaxies. The points are
separated according to the $z$ value. Small open points correspond to  200
ellipticals from  a nearby sample of Virgo, Fornax and Coma clusters. For
clarity error bars are not shown. The mean size relative error is 20\% and the
mean luminosity error 30\%.}

\label{esocomparison}
  \end{figure}

In addition to the  SDSS distribution and for covering the range of faint
galaxies in the relation shown in Fig.~ \ref{relumbt}b, we have used a local
sample of 200 elliptical galaxies from the Virgo, Fornax and Coma cluster (Caon
et al. 1990; 1994; Binggeli \& Jerjen 1998; Guti\'errez et al. 2004). This
sample ranges from -22$<$M$_B$$<$-12. To transform to V--band magnitudes we
have used $B-V$=0.9. To facilitate the comparison with this data, this time we
use the semi--major effective radius. The result of this comparison is shown in
Fig. \ref{esocomparison}.

It is interesting to note that the observed distribution, with a plateau at the
faintest galaxies, is in qualitative good agreement with the distribution of
galaxies found in nearby samples. The apparent absence of evolution in the
shape of the distribution of galaxies in the size--luminosity relation suggests
that the early--type galaxies  were yet assembled  at higher redshift (Fasano
et al. 1998, van Dokkum \& Ellis 2003, Franx et al. 2003, van Dokkum et al.
2003, Daddi et al. 2003, Vazdekis, Trujillo \& Yamada 2004). However, we
observe that at a given size, high--z early--type galaxies present higher
luminosities than the local sample. To quantify this increase we have fitted
the early--type luminosity--size relation of the SDSS by using an expression of
the form: \begin{equation} r_e=a\times L_V^b \end{equation} We find the
following values: a$_{SDSS}$=2.25 and b$_{SDSS}$=0.65. We have repeated the
same exercise this time in our $z$$>$0.5 early--type sample. Due to the
scarcity of points we have fixed the $b$ slope to the same value than the SDSS
sample. To make the fit we use the points defined by the giant ellipticals
(i.e. those points with L$_V$$>$10$^{10}$ h$^{-2}$ L$_\odot$). On doing that we
find: a$_{HDFs}$=1($\pm$0.05). This implies a factor of $\sim$3.5($\pm$0.25)
greater in luminosity which corresponds to $\sim$1.35($\pm$0.10) mag. This is
in good agreement with the brightening of 1--1.5 mag observed in the
rest--frame B--band by Im et al (1996; 2002) and Schade et al. (1999) using the
HST Medium Deep Survey and other HST fields. It is also obtained in  large area
fields (Bell et al. 2004). It is of interest to note that the comparison
between SDSS and the HDFs are between field early--type galaxies. The bright
elliptical galaxies in the nearby clusters define a relation with
a$_{clusters}$=3.45. Consequently, we find that early--type galaxies in the
HDFs at z$\sim$0.7 are a factor of 6.7 brighter in luminosity ($\sim$2 mag)
than the present--day elliptical in clusters. In agreement with previous
studies that indicate that early--types in clusters are older than in the
field.

We have checked, using a 2D KS, test whether the
no--evolution (versus the passive evolution) model can be ruled out. To do that
we have used only the galaxies with L$_V$$>$0.1$\times$10$^{10}$ h$^{-2}$
L$_\odot$ and  z$>$0.5. This test rejects the no--evolution hypothesis with a
confidence larger than than 99.9\%. This luminosity evolution is consistent
with models in which the bulk of stars in the field E/S0 galaxies formed before
$z$$>$1.5 and have been evolving rather quiescently.

\section{Discussion and conclusions}


It is of interest to compare the results presented in this paper with the
theoretical expectations according to the different scenarios of galaxy
formation.

 In the hierarchical scenario, following Mo et al. (1998), the size of the
baryonic component of disc galaxies is expected to scale with redshift as
$R\propto H^{-1}(z)$ at fixed circular velocity, or $R\propto H^{-2/3}(z)$ at
fixed mass. In a flat universe $\Omega$=1, the Hubble constant evolves with
redshift as: \begin{equation} H(z)=H_0[\Omega_\Lambda+\Omega_0(1+z)^3]^{1/2}
\end{equation} where $\Omega_\Lambda$ and $\Omega_0$ are the present--day
density parameters. At $z$$\sim$0.7 (the mean $z$ of our high--z population),
the expected evolution is 0.68$<$R(0.7)/R(0)$<$0.77. This decrease is 
compatible with our estimation: $\sim$30($\pm$10)\%. However, before extracting
any conclusion we must take into account that the Mo et al. predictions are for
the mass--size relation whereas this work deals with the luminosity--size
relation. In fact,  according to previous  (Trujillo et al 2004) and ongoing
(Barden et al. 2004; private communication) work, although the luminosite--size
relation has changed dramatically from z$\sim$3, the stellar mass--size
relation has remained nearly invariant from then.

In the infall scenario, for the rest--frame B--band, Bouwens \& Silk (2002)
propose the following dependence in size: R(z)/R(0)=1-0.27$z$. At $z$=0.7 this
corresponds to a decrease of 19\%. This value is  in agreement (although is
slightly smaller than the observed) with our measurements. Contrary to the
above predictions of the hierarchical models, the Bouwens \& Silk prediction is
for the luminosity--size relation, and consequently, the comparison with our
observations is a more direct one. It is important to note, however, that our
analysis is done in the V--band, and the expected evolution from the  Bouwens
\& Silk (2002) model could be slightly different in this band.

As pointed out by Bouwens \& Silk (2002): ``The hierarchical and infall models
predict relatively similar amounts of evolution in global properties (size,
surface brightness, and luminosity) for disc galaxies to $z$$\sim$1''. This
implies we need to find an extra test for disentangle between both scenarios.
The main difference between both approaches is the evolution of the comoving
number density of the objects. In the infall models is implicitly assumed no
number evolution whereas in the hierarchical model the number of disc galaxies,
is an increasing function of $z$: N(V$_c$,z)dV$_c$ $\propto (1+z)^{(2.9-3.3)}$
at a fixed circular velocity, or N(L,z)dL $\propto (1+z)^{(1.45-1.85)}$, at a
fixed luminosity (see e.g. Mao, Mo \& White 1998). The scarce number of
galaxies in this work and the  strong field--to--field variations prevent a
definitive answer at this point and larger areas are required.

Contrary to the disc galaxies, the number of field E/S0s decreases at high--z
in the hierarchical picture. However, in a flat nonzero $\Lambda$ universe the
semianalytical models predict a weak number density evolution. This makes 
difficult to disentangle  between the monolithic evolution scenario and the
hierarchical view up to $z$$\sim$1. A more definitive answer to this problem
will come from studying in detail the population of E/S0s galaxies in the
1$<$z$<$2 interval.

\section*{Acknowledgments}

We are happy to thank Shiyin Shen for providing us with the Sloan Digital Sky
Survey data used in this paper. We are grateful to Marco Barden, Eric Bell, Ivo
Labb\'e, Daniel H. McIntosh, Hans--Walter Rix and Gregory Rudnick for interest
and useful discussion. We also would like to thank to David Butler for kindly
proof--reading parts of this paper. We wish to thank the anonymous referee for
the helpful comments.

Based on observations made with the NASA/ESA Hubble Space
Telescope, obtained from the data archive at the Space Telescope Institute.
STScI. STScI is operated by the association of Universities for Research in
Astronomy, Inc. under the NASA contract NAS 5--26555.

Funding for the creation and distribution of the SDSS Archive has been provided
by the Alfred P. Sloan Foundation, the Participating Institutions, the National
Aeronautics and Space Administration, the National Science Foundation, the US
Department of Energy, the Japanese Monbukagakusho, and the Max-Planck Society.
The SDSS Web site is http://www.sdss.org. The SDSS is managed by the
Astrophysical Research Consortium (ARC) for the Participating Institutions. The
Participating Institutions are the University of Chicago, Fermilab, the
Institute for Advanced Study, the Japan Participation Group, Johns Hopkins
University, Los Alamos National Laboratory, the Max-Planck-Institut f\"ur
Astronomie (MPIA), the Max-Planck-Institut f\"ur Astrophysik (MPA), New Mexico
State University, University of Pittsburgh, Princeton University, the US Naval
Observatory, and the University of Washington.

\end{document}